\documentclass[12pt]{article}

\newcommand{\blind}{1}

\usepackage{amsmath, amssymb, bm,enumerate, mathrsfs,mathtools, enumitem}
\usepackage{latexsym,color,verbatim, array, multirow, amsthm}
\usepackage{multirow}
\usepackage{array, tabularx, makecell}

\usepackage{etoolbox}
\makeatletter
\patchcmd{\@makecaption}
  {\parbox}
  {\advance\@tempdima-\fontdimen2} 
  {}{}
\makeatother  

\usepackage{enumerate}
\usepackage{comment}
\usepackage{natbib}
\usepackage{caption}
\usepackage{float, bigints}
\usepackage{graphics,amssymb,color}
\usepackage{graphicx}
\usepackage[caption = false]{subfig} 

\usepackage{grffile}
\usepackage{mathtools}

\usepackage{algorithm}
\usepackage{algorithmic}

\newlength\myindent
\usepackage{xfrac}
\usepackage{setspace}

\usepackage{natbib}

\usepackage{hyperref}
\hypersetup{colorlinks,
		citecolor=blue,
		linkcolor=blue,
		urlcolor=black}

\newcommand{\rE}{\mathrm{E}}
\newcommand{\real}{\mathbb{R}}

\newtheorem{proposition}{Proposition}[section]
\newtheorem{theorem}{Theorem}

\newtheorem{remark}{Remark}
\newtheorem{Example}{Example}[section]
\newtheoremstyle{example_contd}
{\topsep} {\topsep}%
{\upshape}
{}
{\bfseries\scshape}
{.}
{1em}
{\thmname{#1} \thmnumber{ #2}\thmnote{#3} (continued)}

\theoremstyle{example_contd}

\newcommand{\grad}{\nabla}

\newcommand*{\Scale}[2][4]{\scalebox{#1}{$#2$}}

\makeatletter
\renewcommand*\env@matrix[1][\arraystretch]{%
  \edef\arraystretch{#1}%
  \hskip -\arraycolsep
  \let\@ifnextchar\new@ifnextchar
  \array{*\c@MaxMatrixCols c}}
\makeatother

\usepackage{makecell}

\newcommand\MyBox[2]{
  \fbox{\lower0.75cm
    \vbox to .6cm{\vfil
      \hbox to 1cm{\hfil\parbox{0.8cm}{#1\\#2}\hfil}
      \vfil}%
  }%
}

\newcolumntype{M}[1]{>{\centering\arraybackslash}m{#1}}
\newcolumntype{N}{@{}m{0pt}@{}}

\begin{document}

\date{}

\def\spacingset#1{\renewcommand{\baselinestretch}%
{#1}\small\normalsize} \spacingset{1}


\if1\blind
{
  \title{\bf Approximate selective inference via maximum likelihood}
  \author{Snigdha Panigrahi \thanks{
    The author acknowledges support by NSF-DMS 1951980 and NSF-DMS 2113342.}\hspace{.2cm} \\
    Department of Statistics,
		University of Michigan,
         MI, USA.\\
    and \\
    Jonathan Taylor \thanks{
    The author acknowledges support in part by ARO grant 70940MA.}\hspace{.2cm} \\
    Department of Statistics,
		 Stanford University,
         CA, USA.}
  \maketitle
} \fi

\if0\blind
{
  \bigskip
  \bigskip
  \bigskip
  \begin{center}
    {\LARGE\bf Approximate selective inference via maximum likelihood}
\end{center}
  \medskip
} \fi

\begin{abstract}
Several strategies have been developed recently to ensure valid inference after model selection; some of these are easy to compute, while others fare better in terms of inferential power.
In this paper, we consider a selective inference framework for Gaussian data. 
We propose a new method for inference through approximate maximum likelihood estimation. 
Our goal is to: (i) achieve better inferential power with the aid of randomization, (ii) bypass expensive MCMC sampling from exact conditional distributions that are hard to evaluate in closed forms.
We construct approximate inference, e.g., p-values, confidence intervals etc., by solving a fairly simple, convex optimization problem. 
We illustrate the potential of our method across wide-ranging values of signal-to-noise ratio in simulations.
On a cancer gene expression data set we find that our method improves upon the inferential power of some commonly used strategies for selective inference.
\end{abstract}

\noindent%
\vfill

\newpage
\spacingset{1.5}

\section{Introduction}
\label{introduction}
\vspace{-2mm}

Querying the data has become a fairly common practice for anyone who wishes to learn a model from a range of different candidates. 
Naively using the same data twice, first to learn a model and then infer for the selected parameters, tends to inflate their estimated effects.
As an example of a query, consider a variable selection algorithm with a shrinkage penalty \citep{tibs_lasso, fan2001variable, yuan2006model}; the algorithm learns a set of variables (or features) into a model.
Ignoring the dependence of the model (and its parameters) on the outcome of the query while calculating p values, confidence intervals, credible intervals etc. undermines inference after selection; see \cite{benjamini2005false, leeb2005model, leeb2006can, berk2013valid} for a demonstration of the concerns here.
The result is usually an increased chance of finding a statistically significant result when the selected variable in fact has no effect.

Various strategies for selective inference offer different solutions by characterizing the dependence between the learned models and data.
Some of the strategies are easier to implement, while others fare better in inferential power.
In this paper, we introduce a new method for selective inference through approximate maximum likelihood estimation.
Building on the recent work by \cite{randomized_response}, our method allows us to: (i) harness Gaussian randomization variables towards better inferential power after selection, and simultaneously (ii) bypass expensive MCMC sampling from intractable conditional distributions. 
Below, we provide a brief, informal overview of our method in a standard setup of linear regression.

\noindent\textbf{An informal overview of our method.}
Consider a regression problem in which we observe a response vector $y\in \real^n$ and a matrix of $p$ predictors $X\in \real^{n \times p}$.
Let $\omega\in \real^p$ be drawn from a centered Gaussian distribution with known covariance $\Sigma_{\mathbb{W}}$.
For fixed values $\lambda \in \real^+$, $\epsilon \in \real^+$, we solve the following query:
\begin{equation}
\label{first:eg}
\underset{o}{\text{minimize}} \; \dfrac{1}{2}\|y- Xo\|_2^2  + \lambda \|o\|_1 + \dfrac{\epsilon}{2} \|o\|_2^2 -\omega^{\intercal} o.
\end{equation}
We call the optimization in \eqref{first:eg} a ``randomized LASSO" query; so named because of the randomization variable $\omega$ added to the objective of the canonical LASSO.
Suppose the query selects a nonempty set of variables $\mathrm{E} \subseteq \{1,2,\cdots, p\}$.
Following selection, we describe our response variable through the model: 
$y = X_{\mathrm{E}}\beta_{\mathrm{E}} + \mathrm{e}, \; \mathrm{e} \sim N(0, \sigma^2 I_n),$
where $I_n$ is the identity matrix with $n$ rows and columns.

A natural ask in the learned model is inference for the partial regression coefficients after adjusting for their dependence on data through $\rE$. 
One concrete course of action is to condition on selection, specifically, base inference on the likelihood of the observed data when conditioned on the event 
$$\left\{ (y, \omega) : \; \widehat{\mathrm{E}}(y, \omega)= \rE\right\},$$
where $\widehat{\mathrm{E}}$ represents the (data-dependent) selected set of variables.
Maximizing the conditional likelihood function gives us $\widehat{\beta}^{\text{\;mle}}_{\rE}$, the maximum likelihood estimate (MLE) for $\beta_{\rE}$.
Taking the hessian of the negative log-likelihood at the MLE yields us $I(\widehat{\beta}_{\rE}^{\text{\;mle}})$, the observed Fisher information matrix.
The (approximate) confidence intervals resulting from our method take the form
\begin{equation}
\label{ci}
\widehat{\beta}^{\text{\;mle}}_{j\cdot \rE} \pm z_{1-q/2} \cdot \sqrt{I^{-1}_{j,j}(\widehat{\beta}_{\rE}^{\text{\;mle}})},
\end{equation}
where $\widehat{\beta}^{\text{\;mle}}_{j\cdot \rE}$ is the $j^{\text{th}}$ component of $\widehat{\beta}^{\text{\;mle}}_{\rE}$, $I^{-1}_{j,j}(\widehat{\beta}_{\rE}^{\text{\;mle}})$ is the $(j,j)^{\text{th}}$ entry of $I^{-1}(\widehat{\beta}_{\rE}^{\text{\;mle}})$ and $z_{1-q}$ is the $(1-q)$-th quantile of a standard normal distribution for $q\in (0,1)$. 

The intervals proposed in \eqref{ci} are seemingly straightforward if only we could directly calculate the two estimates in the expression, the MLE and the observed Fisher information matrix.
But as it turns out, the conditional likelihood and subsequently the two estimates based on it do not admit expressions in closed forms.
In the remaining development, we solve this challenge head-on in two steps. 
First, we construct an approximate, statistically consistent proxy for the likelihood function after conditioning on the selection event. 
Then, we provide a tractable system of estimating equations to obtain the MLE and the observed Fisher information matrix from our proxy likelihood. 
At the core of the proposed estimating equations is a fairly simple, convex optimization problem in relatively few dimensions. 

\noindent\textbf{Comparison with common baselines.}
Continuing with the regression setup, we compare our proposal with some common baselines in a simulated experiment and relate our method with existing work.
In Table \ref{table:eg}, four methods for selective inference including our method are evaluated on three criteria after selecting variables using a fixed value of tuning parameter: average coverage of interval estimates with nominal false coverage rate (FCR) level of $0.10$; average length of the interval estimates; the power of detecting true associations after applying the selective inference strategy.  
The data in this experiment obeys a linear model with a $300$-by-$100$ design matrix $X$ and Gaussian errors, such that the rows of $X$ are i.i.d.~copies of a correlated multivariate normal vector; the model coefficient vector $\beta$ has $6$ nonzero components that are linearly varying in magnitude, and the setting corresponds to a relatively weak signal-to-noise ratio value. 
The simulation setting is described more precisely later in the paper and relative comparisons between all the methods are made for a wide range of values for signal-to-noise ratio.

The first baseline, ``Lee et al.", proposed by \cite{exact_lasso} reduces inference to a truncated normal variable through the Polyhedral Lemma.
The second baseline, ``Split" uses a randomly chosen one-third of the data samples for inference after applying the LASSO to the remaining two-thirds of the data, \cite[e.g.,][]{cox, hurvich1990impact}.
Based on \cite{liu2018more}, the third baseline ``Liu et al." conditions on (strictly) less information than ``Lee et al." by choosing to infer for the parameters associated with the selected variables in the full model: $y \sim N(X\beta, \sigma^2 I_n)$.
All the other strategies in this example use the learned model: $y \sim N(X_{\mathrm{E}}\beta_{\mathrm{E}}, \sigma^2 I_n)$ based on the selected set of variables.

\begin{center}
\captionof{table}{Comparison with baselines.}
\label{table:eg}
\bgroup
\def\arraystretch{1.3}
\scalebox{0.88}{\begin{tabular}{ |c | c  | c  | c | c | }
\hline
 Method  & \makecell{Coverage \\ $100\cdot (1-\text{FCR})\%$} & Lengths  & \makecell{Power}  &  \makecell{$\%$ of infinitely\\  long intervals}\\ [0.5ex] 
\hline \hline
 MLE (Our method)  & $90.92\%$  & $8.31$ & $85\%$  & 0 \\ [0.5ex] 
 \hline
Lee et al. & $85.60\%$  & $\infty$ & $77\%$ & 3.7 \\ [0.5ex] 
\hline
Split & $88.40\%$  & $14.83$ & $56\%$ & 0\\ [0.5ex] 
\hline
Liu et al. & $82.68\%$  & $9.67$ & $64\%$ &  0\\ [0.5ex] 
\hline
\end{tabular}}
\egroup
\end{center}

We note that FCR is (roughly) attained at the nominal level by all the strategies, except ``Liu et al." falls slightly short of the mark in this setting.
Our method, namely ``MLE'', delivers the shortest intervals with the highest power.
In the last column of the table, we also indicate the percentage of intervals that resulted with infinite length. 
Among all the methods, ``Lee et al." produces some infinitely long intervals; this observation is consistent with the established fact in \cite{kivaranovic2018expected} that the intervals based on the Polyhedral Lemma do not have a finite expected value in the Gaussian regression setting.
Both ``Split" and ``Liu et al."  overcome the drawbacks of ``Lee et al." by setting aside more information for inference. 
The former strategy does so by reserving a randomly chosen subsample for inference, while the latter achieves an increase in power through a larger truncation set.

By analogy with \cite{randomized_response}, our method uses added randomization to remedy the excessively long intervals produced by ``Lee et al."; because we do not condition on the randomization variable itself, inference does not trivially reduce to the Polyhedral Lemma.
Our choice of adding a Gaussian randomization variable to the query draws motivation from data carving, a two stage situation where parameters learned on an initial data set are estimated using new samples augmented with the initial ones \citep{optimal_inference, panigrahi2019carving}. 
In the analysis here, the variance of $\omega$ is chosen so that the randomized LASSO (roughly) resembles ``Split" in the amount of information used towards learning the model.
Our method improves upon ``Split" by conditioning upon an event that implies the selection of the variables in the set $\mathrm{E}$ and therefore still consuming some information from the data used in selection.
A direct relation between the power attained with our randomized method and ``Liu et al.", however, is lacking.
Some gain in power reported for the above setting might be attributed to the use of the learned model by our method as opposed to the full model under which ``Liu et al." offers inference.

\smallskip

\noindent{\textbf{Other related work}.} 
Our maximum likelihood method differs from previous proposals in the tools used and the scope of inference.
Existing strategies for selective inference usually require sampling from conditional distributions, either due to the intractability of their exact counterparts or due to the lack of easily available truncation regions.
For example, the pivot described in \cite{randomized_response} lacks exact expressions and is not readily amenable for computational analyses.
The truncation region in \cite{liu2018more} takes a tractable form for the LASSO; however, this form does not directly generalize to other queries.
Other inferential approaches that account for the effects of selection include sampling from a selection-adjusted posterior in \cite{selective_bayesian} and resampling-based approaches such as bootstrap in \cite{mckeague2015adaptive, guo2020inference}.
The computing costs of these approaches are especially acute if two or more queries are applied for learning models, and the construct of valid inference must appropriately take into consideration the effect of each such query.
Bypassing the requirement to sample from intractable conditional distributions after selection, our maximum likelihood method relies on the solution to a simple, convex optimization problem.
Furthermore, this convex problem assumes a separable form under multiple queries which is particularly amenable for parallel computing.
Much of prior work in the area of selective inference has relied on a testing-based approach for real-valued projections of parameters; see for example \cite{yang2016selective, suzumura2017selective, rugamer2018selective}.
In contrast, the scope of the present likelihood-centric approach extends to joint inference for parameter vectors within learned models.

The rest of the paper is organized as follows. 
In Section \ref{univariate:example}, we introduce our approximate proposal in a univariate file drawer problem. 
We describe in Section \ref{convex:queries} our method of selective inference by deriving a system of estimating equations for the MLE and the observed Fisher information matrix after solving a convex query. 
In Section \ref{multiple:queries}, we generalize our prescription to adjust for multiple, convex queries. 
We conduct simulations in Section \ref{simulation} to study the gains with our method over existing baselines. 
We apply our method to gene expression data from The Cancer Genome Atlas in Section \ref{real:data}, corroborating some of the numerical findings in the simulated experiments.
We conclude with a discussion in Section \ref{sec:discussion}.
In the Appendix, we include proofs for our main results and generalize our method of selective inference for multiple convex queries.

\section{MLE Inference: A First Example}
\label{univariate:example}
Before proceeding further, note, we use (i) $\bar{\Phi}(x)$ for the upper tail probability of the standard Gaussian law at $x\in \real$, and (ii) $\phi(u; \nu, \Theta)$ for the (multivariate) Gaussian density function with mean vector $\mu$ and covariance $\Theta$ at the value $u$; the special symbol $\phi(x; 0, 1)$ denotes the standard Gaussian density for $x\in \real$.

\subsection{Univariate soft-truncated likelihood}
\label{uni:file:drawer}

We consider two independent random variables:
$$Y\sim N(\beta, 1), \ W\sim N(0,\eta^2),  $$
 where $W$ denotes a Gaussian randomization variable.
We pursue inference for $\beta$ only if:
 \begin{equation*}
Y+ W>\tau, \ \ \text{ where } \tau=\sqrt{1+\eta^2}\cdot z_{1-q}.
\end{equation*}

We begin by describing a conditional likelihood by conditioning the Gaussian law of $Y$ upon the selection event:
\begin{equation}
\label{threshold:rule}
\left\{ (y,\omega) \in \real^2 :  y+\omega >\tau\right\}. 
\end{equation}
Define $O=Y+ W-\tau$, which we call an optimization variable in our framework. 
Because, $O\lvert Y=y \sim N(y-\tau, \eta^2)$ before conditioning on the event in \eqref{threshold:rule} and the selection event, $y+\omega >\tau$, is equivalent to: $o >0$, the conditional likelihood for $Y$ and $O$ is given by:
\begin{equation*}
\begin{aligned}
& \left(\bar{\Phi}\left(\frac{(\tau-\beta)}{\sqrt{(1+\eta^2)}}\right)\right)^{-1} \phi(Y; \beta, 1)\cdot  \phi(O; Y-\tau, \eta^2) \cdot  1_{(0,\infty)}(O).
\end{aligned} 
\end{equation*} 
Marginalizing over the optimization variable $O$ yields us a likelihood function of $\beta$, that is equal to:
\begin{equation}
\label{lik:randomized:simple:uni}
\begin{aligned}
 \left(\bar{\Phi}\left(\frac{(\tau-\beta)}{{\sqrt{(1+\eta^2)}}}\right)\right)^{-1} \phi(Y; \beta, 1) \cdot \bar{\Phi}\left(\frac{1}{\eta}(\tau -Y)\right).
\end{aligned}
\end{equation} 
Compared to the conditional Gaussian law in the absence of randomization, $Y \;\lvert \;Y>\tau,$  \citep[c.f. Example 2,][]{optimal_inference}, a soft-truncating function replaces the indicator $1_{(\tau,\infty)}(Y)$.
Hereafter, we refer to the resulting function as a ``soft-truncated likelihood".

\subsection{Selective MLE}
\label{uni:selective:mle}

We are now ready to discuss the maximizer of the soft-truncated likelihood in \eqref{lik:randomized:simple:uni} and inspect some properties of this estimate that serve to motivate our approximate pivot in the paper.
Maximizing the log-likelihood gives us the ``selective MLE", $\widehat{\beta}^{\text{\;mle}}$, based on the estimating equation:
\begin{equation}
\label{estimating:MLE}
\grad\alpha( \widehat{\beta}^{\text{\;mle}} )=Y, 
\end{equation}
where $$\alpha(\beta) = \dfrac{1}{2}\beta^2 + \log\bar{\Phi}\left(\frac{(\tau-\beta)}{\sqrt{(1+\eta^2)}}\right).$$
Turning to the distribution of $\widehat{\beta}^{\text{\;mle}}$, we obtain the density for the selective MLE from \eqref{lik:randomized:simple:uni} by applying the simple variable transformation:
$\widehat{\beta}^{\text{\;mle}} =  \grad\alpha^{-1}( Y)$, and note that this density is proportional to:
\begin{equation}
\label{mle:density}
 |\text{det}(\grad^2\alpha(\widehat{\beta}^{\text{\;mle}} ))|\cdot   \phi(\grad\alpha(\widehat{\beta}^{\text{\;mle}}); \beta, 1) \cdot \bar{\Phi}\left(\frac{1}{\eta}(\tau -\grad \alpha(\widehat{\beta}^{\text{\;mle}} ))\right).
\end{equation}

Proposition \ref{uni:pbounded:mle} obtains an upper bound for the mean squared error of the selective MLE. 
An immediate consequence of this bound is a global (asymptotic) consistency guarantee for the selective MLE; i.e., the guarantee continues to hold for the event in \eqref{threshold:rule} even when it has  a vanishing probability as the sample size grows to infinity.
Streamlining the main exposition to focus on finite sample results, we defer the proof for asymptotic consistency to the Appendix C. 
\begin{proposition}
\label{uni:pbounded:mle}
Fix $B=(1+\eta^2)^{-2}{\eta^{4}}$.
Then, we have:
$$\mathbb{E}\left[( \widehat{\beta}^{\text{\;mle}} - \beta)^2 \; \lvert \; Y+W>\tau \right] \leq (B)^{-1}\cdot \text{\normalfont Var}(Y \; \lvert \; Y+W>\tau).
$$
\end{proposition}

\begin{figure}[H]
\begin{center}
\centerline{\includegraphics[height=175pt,width=0.75\linewidth]{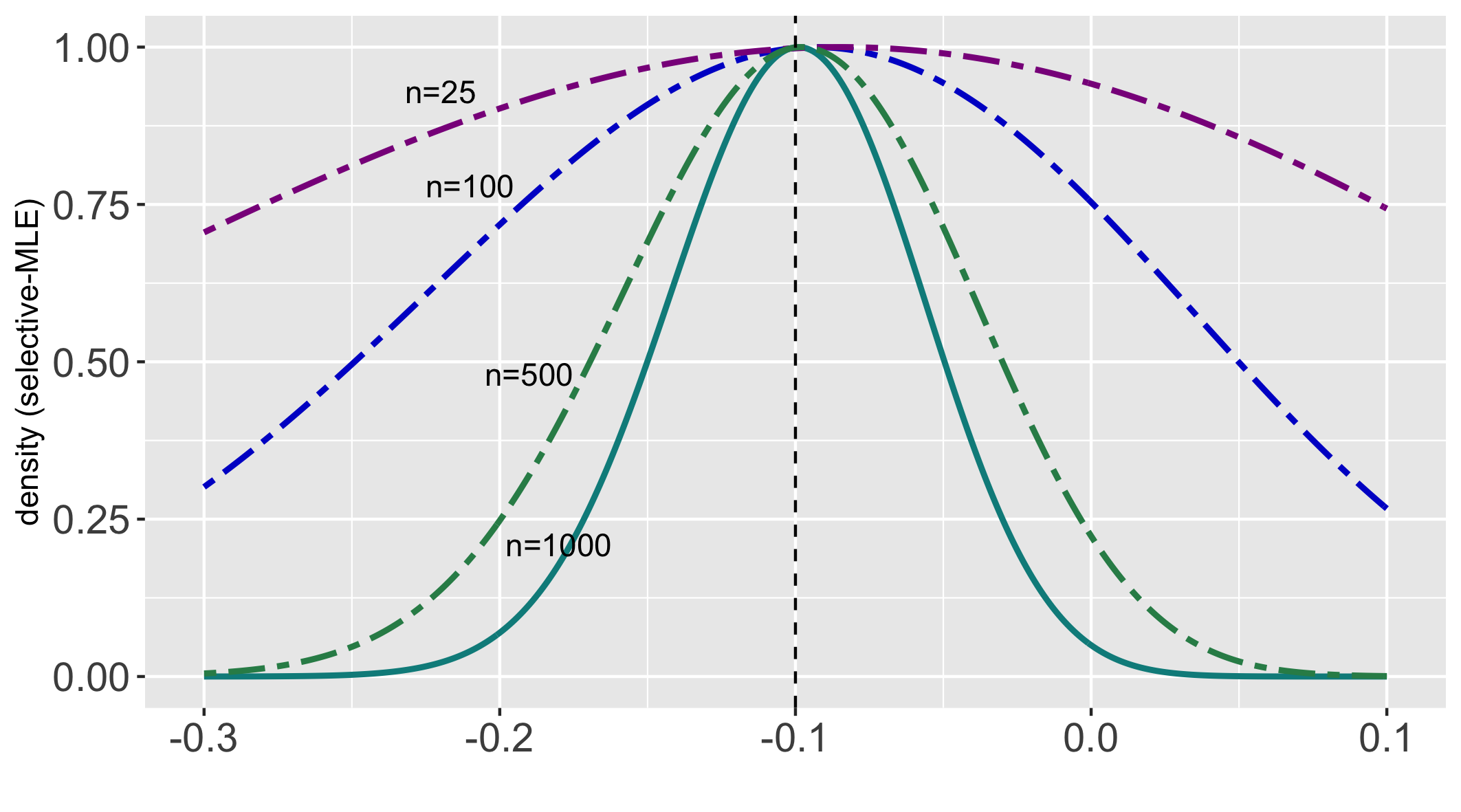}}
\end{center}
 \caption{\small{Distribution of the selective MLE under randomization variance $\eta^2=1$.}}
\label{consistency:0}
\end{figure}

\begin{figure}[H]
\begin{center}
\centerline{\includegraphics[height=175pt,width=0.75\linewidth]{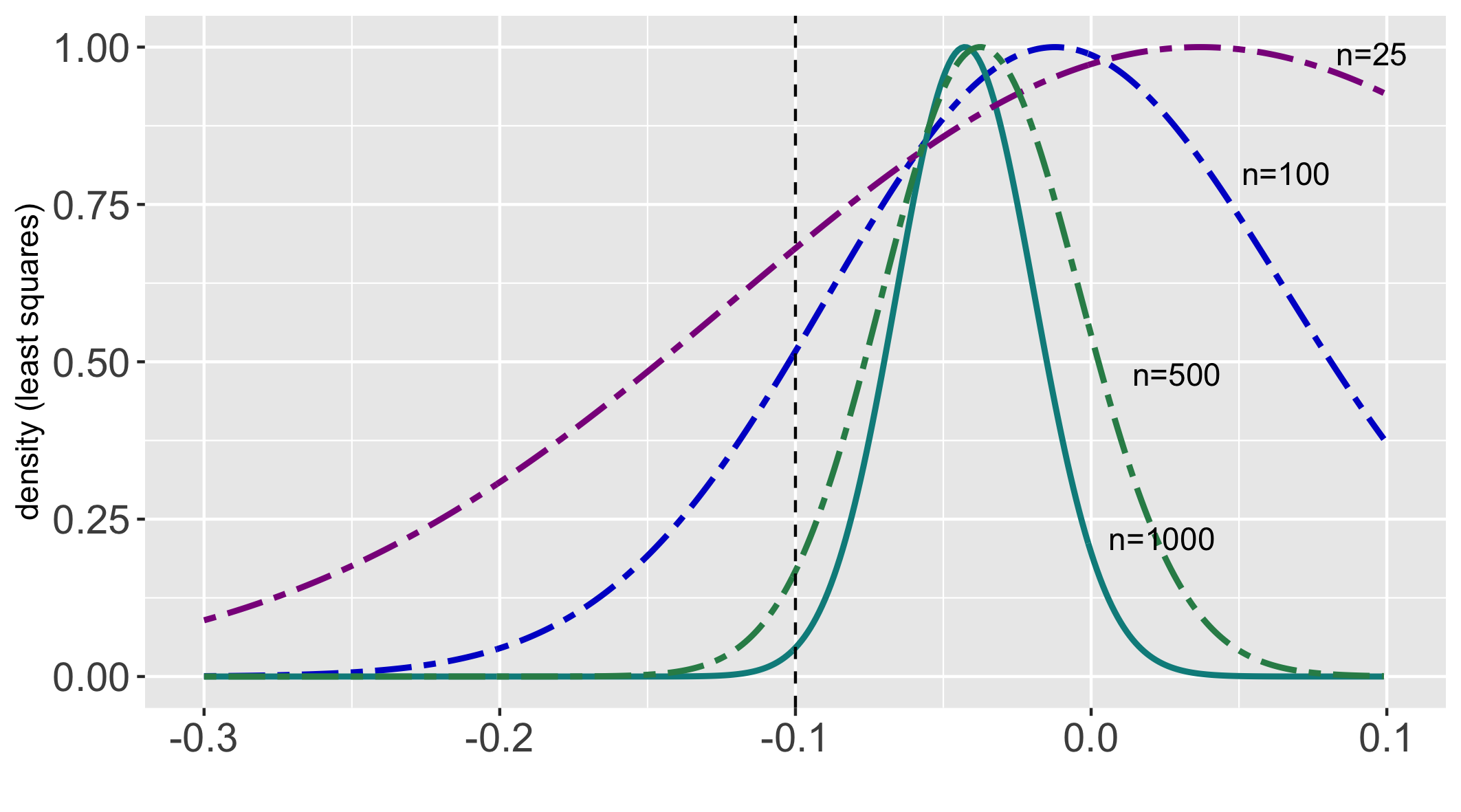}}
\end{center}
 \caption{\small{Distribution of the least squares estimate under randomization variance $\eta^2=1$.}}
\label{consistency:1}
\end{figure}

\begin{figure}[H]
\begin{center}
\centerline{\includegraphics[height=175pt,width=0.75\linewidth]{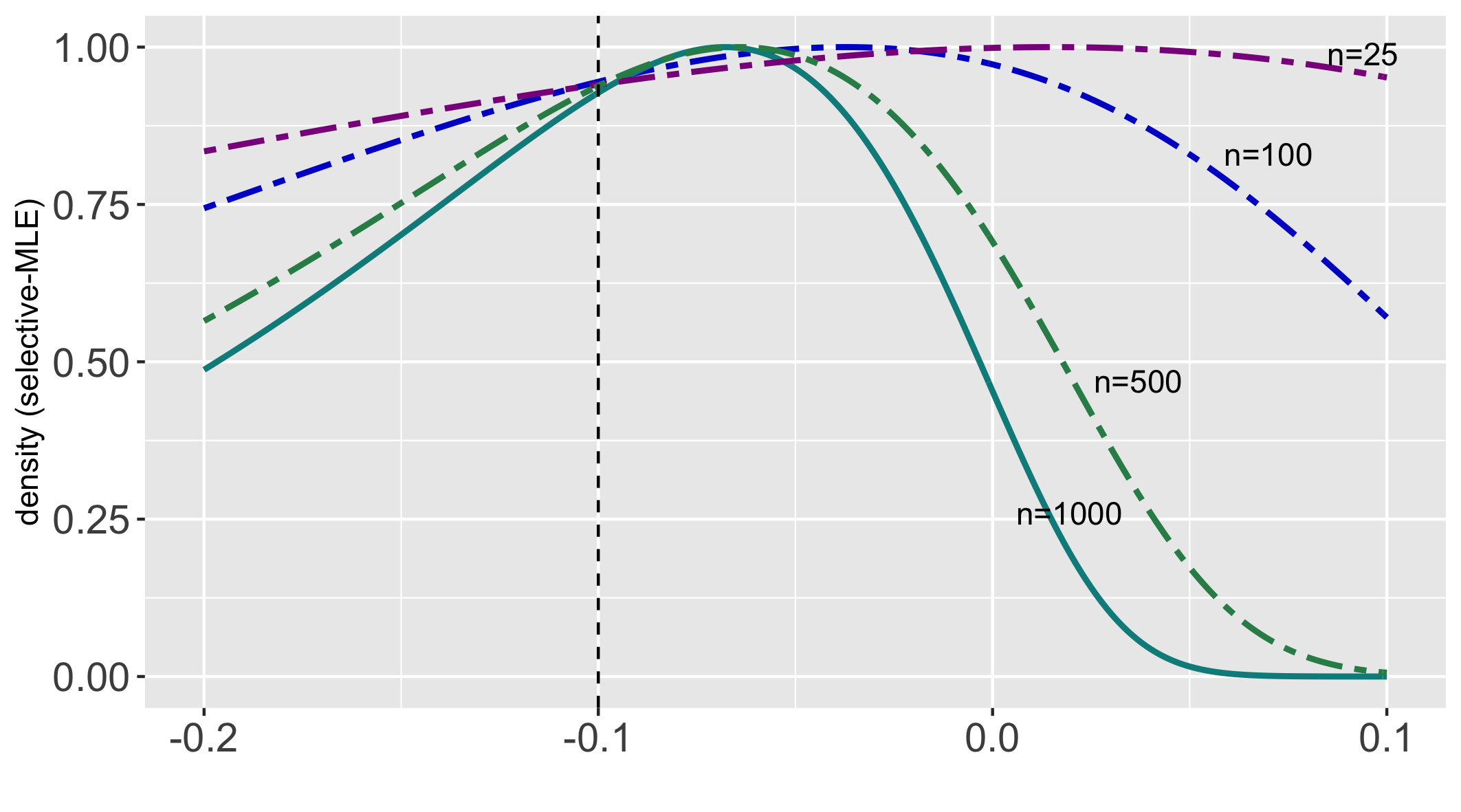}}
\end{center}
 \caption{\small{Distribution of the selective MLE under randomization variance $\eta^2=0.04$.}}
\label{consistency:2}
\end{figure}

Examining for now the asymptotic behavior of selective MLE and the least squares estimate, and the role of the randomization variable $W$, we undertake a simulation by letting $Y:= \sqrt{n}\bar{Y}_n$ with mean $\beta:= \sqrt{n}\beta_n$.
Figures  \ref{consistency:0}, \ref{consistency:1} and \ref{consistency:2} summarize the three primary take-aways from the simulation. 
We let $\tau=0$ in \eqref{threshold:rule}, and fix $\beta_n = \beta_0 =-0.10$ which is highlighted in the figures via a dotted black line.
Notice, our choice of $\beta_n$ results in rarer events of selection with vanishing probabilities as $n\to \infty$.
For the first two figures, the randomization variance $\eta^2$ is equal to $1$.
Based on the density in \eqref{mle:density}, Figure \ref{consistency:0} first studies the behavior of the selective MLE.
Matching our theoretical expectations, the plot demonstrates that the selective MLE is a consistent estimate for the parameter; more specifically, we observe a concentration of the estimate around $\beta_n$ with increasing $n$. 
Next, Figure \ref{consistency:1} replaces the selective MLE in the first plot with the least squares estimate.
Unlike the selective MLE, the least squares estimate fails to concentrate around the parameter of interest for the same sample sizes. 
In Figure \ref{consistency:2}, we  reproduce Figure \ref{consistency:0}, except now we study the behavior of the selective MLE under a very low value of randomization variance, $\eta^2=0.04$. 
An empirical affirmation of the merits of randomization, this plot shows that selective MLE fails to concentrate around $\beta_n$ in the (almost) absence of randomization.

\subsection{Approximate pivot}
 
Prompted by a concentration of the selective MLE around the parameter of interest, we introduce an approximate pivot in the current section.
We propose to approximate the distribution of the MLE by a Gaussian distribution with: (i) mean $\beta$, and (ii) variance equal to inverse of the observed Fisher information, $I(\widehat{\beta}^{\text{\;mle}})$.
Taking a second derivative of the log-likelihood in \eqref{lik:randomized:simple:uni} at the MLE gives us the value of $I(\widehat{\beta}^{\text{\;mle}})$, which is equal to:
\begin{equation}
\label{Fisher-info}
\begin{aligned}
& 1- \dfrac{(\widehat{\beta}^{\text{\;mle}} -\tau)}{(1+\eta^2)^{3/2}}\cdot \left(\bar{\Phi}\left(\frac{(\tau-\widehat{\beta}^{\text{\;mle}} )}{\sqrt{(1+\eta^2)}}\right)\right)^{-1}{
\phi\left(\frac{(\tau-\widehat{\beta}^{\text{\;mle}} )}{\sqrt{(1+\eta^2)}};0,1\right)} \\
&\ \ \ \ \ \ \ \ - \dfrac{1}{(1+\eta^2)}\cdot \left( \bar{\Phi}\left(\frac{(\tau-\widehat{\beta}^{\text{\;mle}} )}{\sqrt{(1+\eta^2)}}\right)\right)^{-2}\phi^2\left(\frac{(\tau-\widehat{\beta}^{\text{\;mle}} )}{\sqrt{(1+\eta^2)}};0,1\right).
 \end{aligned}
\end{equation} 
The Gaussian approximation described above gives rise to the approximate pivot:
\begin{equation}
\label{pivot:approx:uni:filedrawer}
\bar\Phi\left( \sqrt{I(\widehat{\beta}^{\text{\;mle}} )}(\widehat{\beta}^{\text{\;mle}} - \beta)\right).
\end{equation}
We remark that the distribution of the selective MLE, characterized exactly by the density in \eqref{mle:density}, can yield us exact maximum likelihood inference.
In contrast, the pivot in \eqref{pivot:approx:uni:filedrawer} is only approximate in nature, but, appealingly simple in form.
Inference based on the approximate pivot requires us to compute two estimates from the soft-truncated likelihood, namely, the selective MLE and the observed Fisher information.

\begin{figure}[H]
\begin{center}
\centerline{\includegraphics[height=9.cm,width=18.cm]{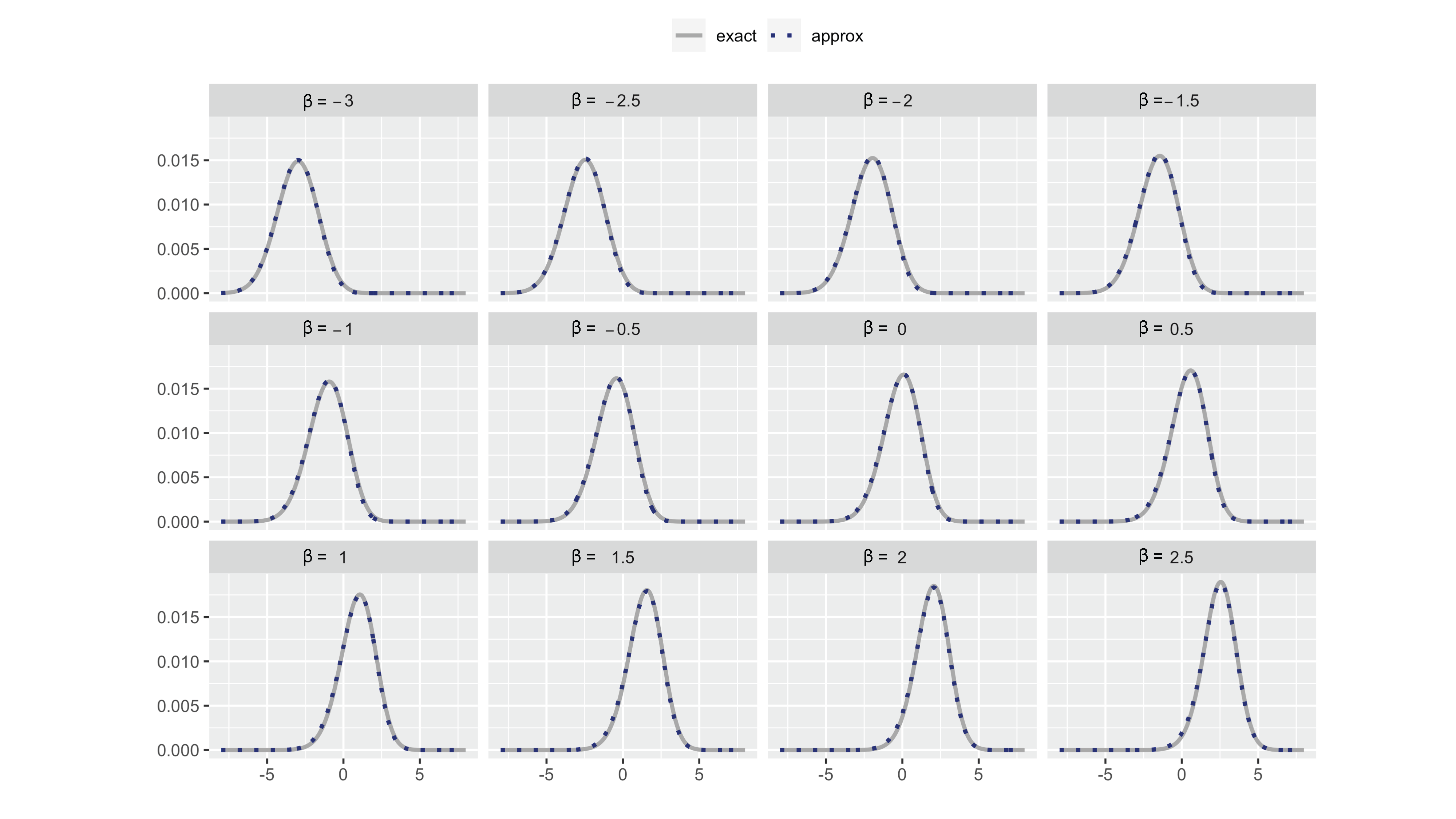}}
\end{center}
\caption{\small{The blue curve represents the normal approximation $N(\beta, I^{-1}(\widehat{\beta}^{\text{\;mle}} ))$ and the gray curve plots the exact density of the MLE in \eqref{mle:density}. }}
\label{density:plot:uni:mle}
\end{figure}

Before turning to the general development, we explore if the proposed Gaussian approximation mimics the exact distribution of the selective MLE. 
In Figure \ref{density:plot:uni:mle}, we represent the density of the selective MLE in \eqref{mle:density}, our benchmark, by the gray curve.  
The panel with $\beta=-3$ results in a rare selection event, while the panel with $\beta=1.5$ results in a highly probable selection event with little selection bias. 
Noteworthy, the effectiveness of our pivot is highlighted via a strong agreement of the proposed (approximate) Gaussian density with the exact (benchmark) density of the selective MLE.

\section{Maximum Likelihood Inference Post Convex Queries}
\label{convex:queries}

We develop our method of maximum likelihood inference below, focusing on the randomized LASSO in a general class of convex queries as our leading example.
In the Appendix, we show that the form of our estimating equations in the primary example generalizes directly to other convex learning queries whose solutions can be similarly characterized through affine Karush-Kuhn-Tucker (K.K.T.) conditions of optimality.

\subsection{Framework under linear regression}
\label{inf:gen:setup}

Consider solving the randomized LASSO in \eqref{first:eg}, where $y$ and $\omega$ denote the observed instances of our response variable $Y\in \real^n$ and randomization variable $W \sim N(0_p, \Sigma_{\mathbb{W}})  \in \real^p$ respectively.
Let $\widehat{\mathrm{E}}(y, \omega) \subseteq \{1,2,\cdots, p\}$ denote the active set of variables selected by the randomized LASSO.
At the solution of the randomized query, we record the value of the subgradient vector for the $\ell_1$ penalty which we represent by $\widehat{\mathrm{S}}(y, \omega)$.
Notice, the collection of instances which lead us to observe $\widehat{\mathrm{S}}(y, \omega) =\mathrm{S}$ result in the active set $\widehat{\mathrm{E}}(y, \omega)=\mathrm{E}$, i.e.,
$$\left\{(y, \omega)  \in\real^n \times \real^p: \widehat{\mathrm{S}}(y, \omega)=\mathrm{S}\right\} \subseteq \left\{(y, \omega)  \in\real^n \times \real^p: \widehat{\mathrm{E}}(y, \omega)=\mathrm{E}\right\}.$$

We turn to a framework for selective inference, allowing our model and parameters to depend on data through the recorded output of our randomized query, $\mathrm{S}$.
Consider a pre-specified mapping $\mathcal{H}: \mathrm{S}\to \mathcal{E}\subseteq \{1,2,\cdots, p\}$.
Our model, after observing $\widehat{\mathrm{S}}(y, \omega)=\mathrm{S}$ and subsequently, noting $\mathcal{E}= \mathcal{H}(\mathrm{S})$, is given by:
 \begin{equation}
\label{post:selective:model}
 \mathbb{M}_{\mathrm{S}} = \Big\{Y\sim N_n(X_{\mathcal{E}}\beta_{\mathcal{E}}, \sigma^2 I), \; \;  \beta_{\mathcal{E}} \in \mathbb{R}^{|\mathcal{E}|} \Big\} \ \text{  for a fixed  }\  \sigma\in \real^{+}.  
 \end{equation}
The mapping $\mathcal{H}$ grants us the flexibility to use (arbitrary) linear models informed by $\mathrm{S}$, including the selected model in the special case when $\mathcal{E}=\mathrm{E}$, the active set of variables.
Suppose, we have a matrix $\mathcal{F}_{\mathrm{S}}\in\mathbb{R}^{d \times n}$, that is allowed to depend on data through $\mathrm{S}$.
Then, let
\begin{equation}
\label{post:selective:target}
\beta_{\mathbb{M}_{\mathrm{S}},\mathrm{S}} = \mathcal{F}_{\mathrm{S}}\mathbb{E}[Y]  \in \real^d
\end{equation}
be our parameter vector of inferential interest. 

In the next step, we form a (multivariate) likelihood function of $\beta_{\mathbb{M}_{\mathrm{S}},\mathrm{S}}$ in the learned model $\mathbb{M}_{\mathrm{S}}$. 
To do so, for a fixed value $\mathrm{S}$, we consider the following statistic: 
$$\widehat{\beta}_{\mathrm{S}} \sim N(\beta_{\mathbb{M}_{\mathrm{S}},\mathrm{S}}, \Sigma_{\mathbb{M}_{\mathrm{S}}, \mathrm{S}}).$$
We derive our soft-truncated likelihood by conditioning the Gaussian law of $\widehat{\beta}_{\mathrm{S}}$ upon the selection event:
\begin{equation}
\label{selection:output}
\left\{(y, \omega)  \in\real^n \times \real^p: \widehat{\mathrm{S}}(y, \omega) =\mathrm{S}\right\}. 
\end{equation} 
The event in \eqref{selection:output} depends not just on $\widehat{\beta}_{\mathrm{S}}$, but further involves a statistic independent of $\widehat{\beta}_{\mathrm{S}}$, which we represent by $\widehat{\beta}^{\perp}_{\mathrm{S}}$.
In addition to the above event, we condition on $\widehat{\beta}^{\perp}_{\mathrm{S}}$ to eliminate nuisance parameters from the likelihood.

For ease of exposition, hereafter, we specialize the above framework to a projected parameter in the selected model
 \begin{equation}
\label{selected:model}
 \Big\{Y \sim N_{n}(X_{\mathrm{E}}\beta_{\mathrm{E}}, \sigma^2 I), \ \beta_{\mathrm{E}} \in \real^{|\mathrm{E}|}\Big\},
 \end{equation}
which we obtain by applying the specific mapping $\mathcal{H}(\mathrm{S})= \mathrm{E}$ and fixing $\mathcal{F}_{\mathrm{S}}= (X_{\mathrm{E}}^{\intercal} X_{\mathrm{E}})^{-1}X_{\mathrm{E}}^{\intercal} \in\mathbb{R}^{|\mathrm{E}| \times n}$.
As noted in \cite{berk2013valid, exact_lasso}, our parameter for inference in this model is the projection of the mean for $Y$ onto the subspace spanned by the columns in $X_{\mathrm{E}}$.
Immediately, we recognize: $\widehat{\beta}_{\mathrm{S}} =(X_{\mathrm{E}}^{\intercal} X_{\mathrm{E}})^{-1}X_{\mathrm{E}}^{\intercal} y$, the least squares statistic refitted to $y$ and $X_{\mathrm{E}}$. 
Besides the least squares statistic, the likelihood involves
 $$
 \widehat{\beta}^{\perp}_{\mathrm{S}} = -X^{\intercal} (y - X_{\mathrm{E}}\widehat{\beta}_{\mathrm{S}}), 
 $$
which is independent of $\widehat{\beta}_{\mathrm{S}}$ under the Gaussian model in \eqref{selected:model}; we will detail this out in the following section.

\subsection{Multivariate soft-truncated likelihood}
\label{gen:lik}

Our main result in the section, Theorem \ref{joint:cond:law}, obtains a soft-truncated likelihood function. 
As seen in the file drawer example, we begin with a compact representation for our selection event in terms of optimization variables based on the randomized LASSO solution.
Introducing some more notations, we let $O_1\in \real^{|\mathrm{E}|}$ and $O_2\in \real^{p-|\mathrm{E}|}$ represent the active (non-zero) components of randomized LASSO solution and the subgradient (sub-)vector for the $\ell_1$ penalty at the inactive indices in $\mathrm{E}^c$ respectively, and let $o_1$ and $o_2$ be the observed instances for these variables.
Let $z_{\mathrm{E}}=\text{sign}(o_1)\in \real^{|\mathrm{E}|}$ be the sign vector for the active components of the estimated LASSO solution.
The K.K.T. conditions for the randomized LASSO are given by:
\begin{equation*} 
 \omega= \begin{pmatrix} \omega^{\intercal}_{\mathrm{E}} & \omega^{\intercal}_{E^c} \end{pmatrix}^{\intercal} = - X^{\intercal} X_{\mathrm{E}}\widehat{\beta}_{\mathrm{S}} +  \begin{bmatrix}X_{\mathrm{E}}^{\intercal} X_{\mathrm{E}} + \epsilon I \\[2pt] X_{\mathrm{E}^c}^{\intercal} X_{\mathrm{E}} \end{bmatrix} o_1 + \begin{pmatrix} \lambda z_{\mathrm{E}} \\ o_2\end{pmatrix} + \widehat{\beta}^{\perp}_{\mathrm{S}}, 
\end{equation*}
where: $-\text{diag}(z_{\mathrm{E}}) \ o_1<0$, and $\|o_2\|_{\infty} <\lambda$.
Because, $\mathrm{S}= \begin{pmatrix}\lambda z^{\intercal}_{\mathrm{E}} &o^{\intercal}_2 \end{pmatrix}^{\intercal},$ our selection event in \eqref{selection:output} is equivalent to the $|\mathrm{E}|$ linear constraints: $U o_1< v$, for the fixed matrices $U= -\text{diag}(z_{\mathrm{E}})$, $v= 0_{|\mathrm{E}|}$.
We note a resemblance with the file drawer example, wherein the selection event is equivalent to the single linear constraint: $o>0$.

Fixing the matrices:
$$P_{\mathrm{S}} = - X^{\intercal} X_{\mathrm{E}}, \; Q_{\mathrm{S}} = \begin{bmatrix}X_{\mathrm{E}}^{\intercal} X_{\mathrm{E}} + \epsilon I \\[2pt] X_{\mathrm{E}^c}^{\intercal} X_{\mathrm{E}} \end{bmatrix},\;r_{\mathrm{S}} = \begin{pmatrix} \lambda z_{\mathrm{E}} \\ o_2\end{pmatrix} + \widehat{\beta}^{\perp}_{\mathrm{S}},$$ 
we rewrite the stationary mapping in the K.K.T. condition as
\begin{equation} 
\label{KKT:gen}
 \omega= \begin{pmatrix} \omega^{\intercal}_{\mathrm{E}} & \omega^{\intercal}_{E^c} \end{pmatrix}^{\intercal} = P_{\mathrm{S}} \widehat{\beta}_{\mathrm{S}} +  Q_{\mathrm{S}} o_1 + r_{\mathrm{S}}.
\end{equation}
Based on
\begin{equation*}
\bar{\Sigma}^{-1} = Q_{\mathrm{S}}^{\intercal}\Sigma_{\mathbb{W}}^{-1}Q_{\mathrm{S}}, \ A= -\bar{\Sigma}Q_{\mathrm{S}}^{\intercal} \Sigma_{\mathbb{W}}^{-1}P_{\mathrm{S}}, \
b=  -\bar{\Sigma}Q_{\mathrm{S}}^{\intercal} \Sigma_{\mathbb{W}}^{-1} r_{\mathrm{S}}, 
\end{equation*}
define:
 \begin{equation}
\label{reference:gen}
f(\widetilde{\beta}_{\mathrm{S}}) =\int \phi(o_1; A\widetilde{\beta}_{\mathrm{S}} +b, \bar{\Sigma})\cdot  1_{\mathrm{R}_0}(o_1) do_1,
\end{equation}
where $\mathrm{R}_0 = \{o_1 \in \real^{|\mathrm{E}|} : Uo_1 < v\}$. 
\begin{theorem}
\label{joint:cond:law} 
After conditioning the law of $\widehat{\beta}_{\mathrm{S}}$ upon $\widehat{\mathrm{S}}(Y, W) =\mathrm{S}$ and $\widehat{\beta}^{\perp}_{\mathrm{S}}(Y)= \widehat{\beta}^{\perp}_{\mathrm{S}}$, the soft-truncated likelihood is
\begin{equation*}
\begin{aligned}
\left(\int \phi(\widetilde{\beta}_{\mathrm{S}}; J\beta_{\mathbb{M}_{\mathrm{S}}, \mathrm{S}} + k, \Sigma) \cdot f(\widetilde{\beta}_{\mathrm{S}})  d\widetilde{\beta}_{\mathrm{S}}\right)^{-1}\phi(\widehat{\beta}_{\mathrm{S}}; J\beta_{\mathbb{M}_{\mathrm{S}}, \mathrm{S}} + k, \Sigma) \cdot  f(\widehat{\beta}_{\mathrm{S}}),
\end{aligned}
\end{equation*}
where $\Sigma$, $J$, $k$ are equal to:
\begin{equation*}
\Sigma = (\Sigma_{\mathbb{M}_{\mathrm{S}}, \mathrm{S}}^{-1} + P_{\mathrm{S}}^{\intercal}\Sigma_{\mathbb{W}}^{-1} P_{\mathrm{S}} -A^{\intercal} \bar{\Sigma}^{-1} A)^{-1}, \ J =\Sigma \Sigma_{\mathbb{M}_{\mathrm{S}}, \mathrm{S}}^{-1}, \ k=\Sigma( A^{\intercal} \bar{\Sigma}^{-1} b- P_{\mathrm{S}}^{\intercal}\Sigma_{\mathbb{W}}^{-1} r_{\mathrm{S}}). 
\end{equation*}
\end{theorem}

\subsection{Approximate inference}

The likelihood in Theorem \ref{joint:cond:law}, though exact, does not directly result in tractable estimating equations for the maximum likelihood estimate and the observed Fisher information matrix.   
This is because the normalizer for the soft-truncated likelihood lacks a closed-form expression.
To circumvent the problem, we propose an approximate proxy for our soft-truncated likelihood based on an upper bound for the normalizer in Proposition \ref{approximation:Gaussianity}. 
Later in the Appendix, using a large deviations principle, we prove that the approximate proxy converges to the exact likelihood with increasing sample size.
Furthermore, we show the maximizer of the approximate likelihood, $\widehat{\beta}^{\;\text{mle}}_{\mathbb{M}_{\mathrm{S}},\mathrm{S}}$, is guaranteed to concentrate around the parameter in \eqref{post:selective:target}. 
Endowed with a property we expect with the exact MLE, we call the maximizer of the approximate likelihood an ``approximate selective MLE".

\begin{proposition}
\label{approximation:Gaussianity}
Let $\mathrm{R}$ be a convex and compact subset of $\real^{|\mathrm{E}|}\times \real^{|\mathrm{E}|}$. 
Suppose, $\widehat{\beta}_{\mathrm{S}}$ and $O_1$ are drawn from a Gaussian distribution with the following likelihood: 
\begin{equation*}
\label{joint:Gaussian:den}
\phi(\widehat{\beta}_{\mathrm{S}}; J\beta_{\mathbb{M}_{\mathrm{\mathrm{S}}}, \mathrm{\mathrm{S}}} + k, \Sigma)\cdot \phi(O_1; A \widehat{\beta}_{\mathrm{S}} + b, \bar{\Sigma}). 
\end{equation*}
Then, $\log\mathbb{P}\left[\begin{pmatrix}\widehat{\beta}_{\mathrm{S}}^\intercal , O_1^\intercal \end{pmatrix}^\intercal \in \mathrm{R}\right]$ is bounded from above by
\vspace{-4mm}
\begin{equation*}
\begin{aligned}
 - \displaystyle\inf_{(\widetilde{\beta}_{\mathrm{S}}, o_1)\in \mathrm{R}}\Big\{\dfrac{1}{2} (\widetilde{\beta}_{\mathrm{S}}-  J\beta_{\mathbb{M}_{\mathrm{\mathrm{S}}}, \mathrm{\mathrm{S}}} - k)^{\intercal} & \Sigma^{-1}(\widetilde{\beta}_{\mathrm{S}}- J\beta_{\mathbb{M}_{\mathrm{\mathrm{S}}}, \mathrm{\mathrm{S}}} - k)\\
&+  \dfrac{1}{2} (o_1  -A\widetilde{\beta}_{\mathrm{S}} -b)^{\intercal} \bar{\Sigma}^{-1}(o_1-A\widetilde{\beta}_{\mathrm{S}} -b)\Big\}.
\end{aligned}
\end{equation*} 
\end{proposition}
\vspace{-2mm}

Recall, $\mathrm{R}_0 = \{o_1 \in \real^{|\mathrm{E}|} : Uo_1 < v\}$. 
We apply the bound in Proposition \ref{approximation:Gaussianity} to obtain the following proxy for the exact log-likelihood:
\begin{align*}
\log \phi(\widehat{\beta}_{\mathrm{S}}; J\beta_{\mathbb{M}_{\mathrm{\mathrm{S}}}, \mathrm{\mathrm{S}}} + k, \Sigma)  + \displaystyle\inf_{(\widetilde{\beta}_{\mathrm{S}}, o_1)\in \real^{|\mathrm{E}|} \times \mathrm{R}_0}\; &\Big\{ \dfrac{1}{2} (\widetilde{\beta}_{\mathrm{S}}- J\beta_{\mathbb{M}_{\mathrm{\mathrm{S}}}, \mathrm{\mathrm{S}}} - k)^{\intercal} \Sigma^{-1}(\widetilde{\beta}_{\mathrm{S}}- J\beta_{\mathbb{M}_{\mathrm{\mathrm{S}}}, \mathrm{\mathrm{S}}} - k) \\
&+  \dfrac{1}{2} (o_1  -A\widetilde{\beta}_{\mathrm{S}} -b)^{\intercal} \bar{\Sigma}^{-1}(o_1-A\widetilde{\beta}_{\mathrm{S}} -b) \Big\},
\end{align*}
after ignoring constants free of the parameter vector.
\begin{remark}
Observe, $\real^{|\mathrm{E}|}\times \mathrm{R}_0$, the subset of $\;\real^{|\mathrm{E}|}\times \real^{|\mathrm{E}|}$ associated with the our selection event is clearly not compact. 
While compactness is a requirement to prove that the approximation in Proposition \ref{approximation:Gaussianity} is an upper bound for the normalizer of the likelihood, in practice, we may consider a sufficiently large compact, convex subset such that the probability of the associated event converges to the actual probability with increasing sample size. 
\end{remark}
As noted in \cite{panigrahi2018scalable}, we can further modify the approximation in Proposition \ref{approximation:Gaussianity} to solve an unconstrained optimization via a barrier penalty that reflects the same constraints, but allocates a higher preference to the optimizing variables within the selection region. 
 Letting $\mathcal{B}_{U; v}(o_1)$ denote a barrier penalty for the constraints $Uo_1<v$, the final expression for our approximate log-likelihood agrees up to an additive constant with:
\begin{equation}
\label{barrier:approx}
\begin{aligned}
& \log \phi(\widehat{\beta}_{\mathrm{S}}; J\beta_{\mathbb{M}_{\mathrm{\mathrm{S}}}, \mathrm{\mathrm{S}}} + k, \Sigma)  + \displaystyle\inf_{(\widetilde{\beta}_{\mathrm{S}}, o_1)}\; \Big\{ \dfrac{1}{2} (\widetilde{\beta}_{\mathrm{S}}- J\beta_{\mathbb{M}_{\mathrm{\mathrm{S}}}, \mathrm{\mathrm{S}}} - k)^{\intercal} \Sigma^{-1}(\widetilde{\beta}_{\mathrm{S}}- J\beta_{\mathbb{M}_{\mathrm{\mathrm{S}}}, \mathrm{\mathrm{S}}} - k) \\
&\;\;\;\;\; \;\;\;\;\; \;\;\;\;\; \;\;\;\;\; \;\;\;\;\; \;\;\;\;\; \;\;\;\;\; \;\; \;\;\;\;\;\;\;\;+  \dfrac{1}{2} (o_1  -A\widetilde{\beta}_{\mathrm{S}} -b)^{\intercal} \bar{\Sigma}^{-1}(o_1-A\widetilde{\beta}_{\mathrm{S}} -b) + \mathcal{B}_{U; v}(o_1)\Big\}.
\end{aligned}
\end{equation}
Based on the approximate likelihood in \eqref{barrier:approx}, the results in Theorem \ref{approximate:selective:mle} and Theorem \ref{approximate:Fisher:information} give us compact estimating equations for the two ingredients of approximate maximum likelihood inference.

\begin{theorem}
\label{approximate:selective:mle}
Consider the optimization problem
\begin{equation}
\label{core:optimizer}
o_1^*(\widehat{\beta}_{\mathrm{S}}) =\underset{o_1}{\text{argmin}}\  \dfrac{1}{2} (o_1- A\widehat{\beta}_{\mathrm{S}} -b)^\intercal \bar{\Sigma}^{-1}(o_1- A\widehat{\beta}_{\mathrm{S}} -b) +\mathcal{B}_{U; v}(o_1).
\end{equation}
Then, maximizing the approximate log-likelihood in \eqref{barrier:approx} yields us the following estimating equation for the approximate selective MLE:
\[\widehat{\beta}^{\;\text{mle}}_{\mathbb{M}_{\mathrm{S}},\mathrm{S}}  =  J^{-1}\widehat{\beta}_{\mathrm{S}} - J^{-1}k+\Sigma_{\mathbb{M}_{\mathrm{S}}, \mathrm{S}}  A^\intercal \bar{\Sigma}^{-1} (A \widehat{\beta}_{\mathrm{S}} + b -o_1^*(\widehat{\beta}_{\mathrm{S}})).
\]
\end{theorem}
 
In line with Proposition \ref{uni:pbounded:mle} for the file drawer example, Theorem \ref{multi:pbounded:mle} provides a bound for the mean squared error of the approximate selective MLE.
The bound in this result allows us to formalize a global consistency guarantee for our estimate in Appendix C. 
\begin{theorem}
\label{multi:pbounded:mle}
Let the smallest eigen values for $(\Sigma^{-1}_{\mathbb{M}_{\mathrm{S}}, \mathrm{S}}  + P_{\mathrm{S}}^\intercal \Sigma_{\mathbb{W}}^{-1}P_{\mathrm{S}})^{-1}$ and $\Sigma^{-1}_{\mathbb{M}_{\mathrm{S}}, \mathrm{S}}$
be $\lambda_{0}$ and $\lambda_1$ respectively. 
Fix $B= (\lambda_0\cdot \lambda_1)^{2}$.
Based on the real-valued mapping:
\begin{equation*}
\begin{aligned}
\alpha(\eta_{\mathbb{M}_{\mathrm{S}}, \mathrm{S}})&=\frac{1}{2}\eta_{\mathbb{M}_{\mathrm{S}}, \mathrm{S}}^\intercal \Sigma \eta_{\mathbb{M}_{\mathrm{S}}, \mathrm{S}} - \displaystyle\inf_{(\widetilde{\beta}_{\mathrm{S}}, o_1)\in \real^{|\mathrm{E}|} \times \mathrm{R}_0}\; \Big\{ \dfrac{1}{2} (\widetilde{\beta}_{\mathrm{S}}- \Sigma \eta_{\mathbb{M}_{\mathrm{S}}, \mathrm{S}})^{\intercal} \Sigma^{-1}(\widetilde{\beta}_{\mathrm{S}}- \Sigma \eta_{\mathbb{M}_{\mathrm{S}}, \mathrm{S}}) \\
&\;\;\;\;\; \;\;\;\;\; \;\;\;\;\; \;\;\;\;\; \;\;\;\;\; \;\;\;\;\; \;\;\;\;\; \;\;\;+  \dfrac{1}{2} (o_1  -A\widetilde{\beta}_{\mathrm{S}} -b)^{\intercal} \bar{\Sigma}^{-1}(o_1-A\widetilde{\beta}_{\mathrm{S}} -b) + \mathcal{B}_{U; v}(o_1)\Big\},
\end{aligned}
\end{equation*}
we have
\begin{equation*}
\begin{aligned}
&\mathbb{E}\left[\|\widehat{\beta}^{\;\text{mle}}_{\mathbb{M}_{\mathrm{S}},\mathrm{S}} - \beta_{\mathbb{M}_{\mathrm{S}}, \mathrm{S}}\|_2^2 \; \lvert \; \widehat{\mathrm{S}}(Y, W) =\mathrm{S}, \ \widehat{\beta}^{\perp}_{\mathrm{S}}(Y)= \widehat{\beta}^{\perp}_{\mathrm{S}} \right]\\ 
&\;\;\;\;\;\;\;\;\;\;\;\;\;\;\;\leq (B)^{-1}\mathbb{E}\left[\|\widehat{\beta}_{\mathrm{S}}- \grad \alpha( \Sigma^{-1}(J \beta_{\mathbb{M}_{\mathrm{S}}, \mathrm{S}} + k))\|_2^2 \; \lvert \; \widehat{\mathrm{S}}(Y, W) =\mathrm{S}, \ \widehat{\beta}^{\perp}_{\mathrm{S}}(Y)= \widehat{\beta}^{\perp}_{\mathrm{S}}\right].
\end{aligned}
\end{equation*}
\end{theorem}

We provide a proxy for the observed Fisher information matrix based on the estimate in Theorem \ref{approximate:selective:mle}.
\begin{theorem}
\label{approximate:Fisher:information}
Let $o_1^*(\widehat{\beta}_{\mathrm{S}})$ be the solution to the optimization problem in \eqref{core:optimizer}.
The observed Fisher information $I(\widehat{\beta}^{\;\text{mle}}_{\mathbb{M}_{\mathrm{S}},\mathrm{S}})$ for the approximate log-likelihood in \eqref{barrier:approx} is:
\vspace{-4mm}
\[\Sigma^{-1}_{\mathbb{M}_{\mathrm{S}}, \mathrm{S}} \left(\Sigma^{-1} + A^\intercal \bar{\Sigma}^{-1} A -  A^\intercal \bar{\Sigma}^{-1} (\bar{\Sigma}^{-1}+ \grad^2 \mathcal{B}_{U; v}(o_1^*(\widehat{\beta}_{\mathrm{S}})))^{-1} \bar{\Sigma}^{-1}A\right)^{-1} \Sigma^{-1}_{\mathbb{M}_{\mathrm{S}}, \mathrm{S}}.\vspace{-2mm}\] 
\end{theorem}

We summarize in Algorithm \ref{alg:conf:int} our steps for maximum likelihood inference. 
Our primary computational step is the simple, $\real^{|\mathrm{E}|}$-dimensional, convex optimization problem $\textbf{(O)}$ resulting in \textbf{(S-MLE)} and \textbf{(FI)}.
Emphasized earlier, the form of the estimating equations for the randomized LASSO generalizes to convex queries with affine K.K.T. conditions of optimality as in \eqref{KKT:gen}.
In Appendix B, we illustrate how our method applies to: (i) variable screening based on marginal correlations \citep{lee2014exact}; (ii) variable selection via SLOPE \citep{bogdan2015slope}.

\begin{algorithm}[h]
 \caption{ALGORITHM 1: Approximate maximum likelihood inference post a convex query}
 \label{alg:conf:int}
\begin{spacing}{1.8}
\begin{algorithmic} 
\REQUIRE Query, $\omega\sim N(0, \Sigma_{\mathbb{W}})$
\ENSURE $\widehat{\mathrm{S}}= \mathrm{S}$
\STATE {\text{Implied parameters \textbf{(P)}}}: Compute matrices: $\bar{\Sigma}$, $A$, $b$, $\Sigma$, $J$, $k$
\STATE {\text{Optimization \textbf{(O)}}}: $o_1^*( \widehat{\beta}_{\mathrm{S}} ) = \underset{o_1}{\text{argmin}}\dfrac{1}{2}(o_1- A\widehat{\beta}_{\mathrm{S}}  -b)^T\bar{\Sigma}^{-1}(o_1- A\widehat{\beta}_{\mathrm{S}}  -b)+\mathcal{B}_{U; v}(o_1).$
\STATE{\text{Selective MLE \textbf{(S-MLE)}}}: $\widehat{\beta}^{\;\text{mle}}_{\mathbb{M}_{\mathrm{S}},\mathrm{S}}  =  J^{-1}\widehat{\beta}_{\mathrm{S}} - J^{-1}k+\Sigma_{\mathbb{M}_{\mathrm{S}}, \mathrm{S}}  A^\intercal \bar{\Sigma}^{-1} (A \widehat{\beta}_{\mathrm{S}} + b -o_1^*(\widehat{\beta}_{\mathrm{S}}))$
\STATE{\text{Inverse info \textbf{(FI)}}}: $I^{-1}(\widehat{\beta}^{\;\text{mle}}_{\mathbb{M}_{\mathrm{S}},\mathrm{S}} )=\Sigma_{\mathbb{M}_{\mathrm{S}}, \mathrm{S}} \Big(\Sigma^{-1} + A^T \bar{\Sigma}^{-1} A$ \\ \hspace{65mm} $-A^T \bar{\Sigma}^{-1} (\bar{\Sigma}^{-1}+ \grad^2 \mathcal{B}_{K}(o_1^*( \widehat{\beta}_{\mathrm{S}} )))^{-1} \bar{\Sigma}^{-1}A\Big) \Sigma_{\mathbb{M}_{\mathrm{S}}, \mathrm{S}}$

\STATE{\text{MLE-based inference}}:
\FORALL {$j$ in selected set $\mathrm{E}$}
\STATE{\bf{(\text{p-value for }$\beta_{j; \mathbb{M}_{\mathrm{S}}, \mathrm{S}}$)} }: $2 \min\left(\bar{\Phi}\left(\widehat{\beta}_{j;\mathbb{M}_{\mathrm{S}},\mathrm{S}}^{\;\text{\;mle}}/\sqrt{I_{j,j}^{-1}(\widehat{\beta}^{\;\text{mle}}_{\mathbb{M}_{\mathrm{S}},\mathrm{S}} )}\right), \Phi\left(\widehat{\beta}_{j;\mathbb{M}_{\mathrm{S}},\mathrm{S}}^{\;\text{\;mle}}/\sqrt{I_{j,j}^{-1}(\widehat{\beta}^{\;\text{mle}}_{\mathbb{M}_{\mathrm{S}},\mathrm{S}} )}\right) \right)$
\STATE{\bf{(\text{interval for} $\beta_{j; \mathbb{M}_{\mathrm{S}}, \mathrm{S}}$)} }: $\left(\widehat{\beta}_{j;\mathbb{M}_{\mathrm{S}}, \mathrm{S}}^{\;\text{\;mle}}- z_{1-q/2}\cdot \sqrt{I_{j,j}^{-1}(\widehat{\beta}^{\;\text{mle}}_{\mathbb{M}_{\mathrm{S}},\mathrm{S}} )}, \widehat{\beta}_{j;\mathbb{M}_{\mathrm{S}},\mathrm{S}}^{\;\text{\;mle}}+ z_{1-q/2}\cdot \sqrt{I_{j,j}^{-1}(\widehat{\beta}^{\;\text{mle}}_{\mathbb{M}_{\mathrm{S}},\mathrm{S}} )}\right)$
\ENDFOR
\end{algorithmic}
\end{spacing}
\end{algorithm}

 \section{Maximum Likelihood Inference post Multiple Queries}
\label{multiple:queries}

We now turn to the case of multiple convex queries, outlining our method after $L$ randomized LASSO queries. 
The randomized LASSO query yet again serves only as our primary example for ease of presentation.
More generally, any of the LASSO queries can be replaced with a convex query that admits affine K.K.T. conditions of optimality at the solution.  

Let $\widehat{\mathrm{S}}^{(l)}$ be the subgradient vector for the $\ell_1$ penalty at the solution of the $l^{\text{th}}$ randomized LASSO query:
\begin{equation}
\label{l:LASSO}
\underset{o}{\text{minimize}} \; \dfrac{1}{2}\|y- X o\|_2^2  + \lambda \|o\|_1 + \dfrac{\epsilon}{2} \|o\|_2^2 -(\omega^{(l)})^{\intercal} o, \ \omega^{(l)}  \stackrel{\text{i.i.d.}}{\sim} N_p(0, \Sigma_{\mathbb{W}})
\end{equation}
for $l\in\{1,2,\cdots, L\}$, and let 
$$\widehat{\mathrm{S}}=\begin{pmatrix}(\widehat{\mathrm{S}}^{(1)})^{\intercal} & \cdots &  (\widehat{\mathrm{S}}^{(L)})^{\intercal} \end{pmatrix}^{\intercal}.$$
Consider the event in Equation 11.
Theorem \ref{separability:multi} and \ref{Fisher:info:multi} give the estimating equations for the approximate selective MLE and the observed Fisher information matrix after accounting for the effect of each query on inference. 
Notably, the system of estimating equations relies on $L$ separable optimization problems that can be solved in parallel.

Invoking the framework for selective inference in the preceding section, suppose, we apply the mapping $\mathcal{H}(\mathrm{S})= \mathrm{E}= \cup_{l=1}^L \mathrm{E}^{(l)}$ and fix $\mathcal{F}_{\mathrm{S}}= (X_{\mathrm{E}}^{\intercal} X_{\mathrm{E}})^{-1}X_{\mathrm{E}}^{\intercal}$.
We let $o_1^{(l)}$ and $o_2^{(l)}$ be the realized instances for the active components of the randomized LASSO solution and the subgradient (sub-)vector of the $\ell_1$ penalty at the inactive indices in $\mathrm{E}^c$, respectively. 
Let $z_{\mathrm{E}}^{(l)} = \text{sign}(o_1^{(l)})$.
The K.K.T. conditions of optimality for each randomized LASSO query are given by:
\begin{equation} 
\label{KKT:gen}
 \omega^{(l)}= \begin{pmatrix} (\omega^{(l)}_{\mathrm{E}})^{\intercal} & (\omega^{(l)}_{E^c})^{\intercal} \end{pmatrix}^{\intercal} = P_{\mathrm{S}} \widehat{\beta}_{\mathrm{S}} +  Q^{(l)}_{\mathrm{S}} o^{(l)}_1 + r^{(l)}_{\mathrm{S}}(\widehat{\beta}^{\perp}_{\mathrm{S}}; o^{(l)}_2), 
\end{equation}
together with the constraints: $-\text{diag}(z_{\mathrm{E}}) \ o^{(l)}_1<0$, and $\|o^{(l)}_2\|_{\infty} <\lambda$, where $\widehat{\beta}_{\mathrm{S}} =(X_{\mathrm{E}}^{\intercal} X_{\mathrm{E}})^{-1}X_{\mathrm{E}}^{\intercal} y$, $\widehat{\beta}^{\perp}_{\mathrm{S}} = -X^{\intercal} (y - X_{\mathrm{E}}\widehat{\beta}_{\mathrm{S}}),$
and
$$P_{\mathrm{S}} = - X^{\intercal} X_{\mathrm{E}}, \; Q_{\mathrm{S}} = \begin{bmatrix}X_{\mathrm{E}^{(l)}}^{\intercal} X_{\mathrm{E}^{(l)}} + \epsilon I \\[2pt] X_{(\mathrm{E}^{(l)})^c}^{\intercal} X_{\mathrm{E}^{(l)}} \end{bmatrix},\;r_{\mathrm{S}} = \begin{pmatrix} \lambda z^{(l)}_{\mathrm{E}} \\ o^{(l)}_2\end{pmatrix} + \widehat{\beta}^{\perp}_{\mathrm{S}}. $$ 
Our selection event in Equation 11 is equivalent to $L\cdot |\mathrm{E}|$ linear constraints: $U^{(l)} o^{(l)}_1< v^{(l)}$ for $l\in \{1,2,\cdots, L\}$, where  $U^{(l)}= -\text{diag}(z_{\mathrm{E}^{(l)}})$ and  $v^{(l)}= 0_{|\mathrm{E}|}$.

Using the matrices
\begin{equation*}
 \bar{\Sigma}^{(l)}=\left((Q_{\mathrm{S}}^{(l)})^{\intercal}\Sigma_{\mathbb{W}}^{-1}Q^{(l)}_{\mathrm{S}}\right)^{-1},
 A^{(l)}= -\bar{\Sigma}^{(l)} (Q_{\mathrm{S}}^{(l)})^{\intercal} \Sigma_{\mathbb{W}}^{-1} P^{(l)}_{\mathrm{S}}, 
 b^{(l)}=-\bar{\Sigma}^{(l)}(Q_{\mathrm{S}}^{(l)})^{\intercal} \Sigma_{\mathbb{W}}^{-1} r^{(l)}_{\mathrm{S}},
 \end{equation*}
\begin{equation*}
\begin{aligned}
& \Sigma = \Big(\Sigma_{\mathbb{M}_{\mathrm{S}}, \mathrm{S}}^{-1} + \textstyle\sum_{l=1}^L  \left\{ (P^{(l)}_{\mathrm{S}})^T\Sigma_{\mathbb{W}}^{-1} P^{(l)}_{\mathrm{S}} - (A^{(l)})^T (\bar{\Sigma}^{(l)})^{-1} A^{(l)}\right\}\Big)^{-1}, \\
& J = \Sigma_{\mathrm{S}}\Sigma_{\mathbb{M}_{\mathrm{S}}, \mathrm{S}}^{-1},\;  k= \Sigma_S\Big(\sum_{l=1}^L  \left\{ (A^{(l)})^T (\bar{\Sigma}^{(l)})^{-1} b^{(l)}- (P^{(l)}_{\mathrm{S}})^T\Sigma_{\mathbb{W}}^{-1} r^{(l)}_{\mathrm{S}} \right\}\Big),
\end{aligned}
\end{equation*}
and ignoring an additive constant, an approximate proxy for the exact log-likelihood agrees with
\begin{equation}
\label{barrier:approx:multi}
\begin{aligned}
& \log \phi(\widehat{\beta}_{\mathrm{S}}; J\beta_{\mathbb{M}_{\mathrm{\mathrm{S}}}, \mathrm{\mathrm{S}}} + k, \Sigma)  + \displaystyle\inf_{\widetilde{\beta}_{\mathrm{S}}, o^{(l)}_1,\; l\in \{1,2,\cdots, L\}}\; \Big\{ \dfrac{1}{2} (\widetilde{\beta}_{\mathrm{S}}- J\beta_{\mathbb{M}_{\mathrm{\mathrm{S}}}, \mathrm{\mathrm{S}}} - k)^{\intercal} \Sigma^{-1}(\widetilde{\beta}_{\mathrm{S}}- J\beta_{\mathbb{M}_{\mathrm{\mathrm{S}}}, \mathrm{\mathrm{S}}} - k) \\
&\;\;\;\; +  \displaystyle\sum_{l=1}^L \dfrac{1}{2} (o^{(l)}_1  -A^{(l)}\widetilde{\beta}_{\mathrm{S}} -b^{(l)})^{\intercal} (\bar{\Sigma}^{(l)})^{-1}(o^{(l)}_1-A^{(l)}\widetilde{\beta}_{\mathrm{S}} -b^{(l)})+  +  \displaystyle\sum_{l=1}^L \mathcal{B}_{U^{(l)}; v^{(l)}}(o^{(l)}_1)\Big\}.
\end{aligned}
\end{equation}
In line with Section 3 of the paper, the approximate proxy is motivated by an upper bound for the normalizer of the soft-truncated likelihood, given in the following Proposition.

\begin{proposition}
\label{approximation:Gaussianity:multi}
Suppose, $\widehat{\beta}_{\mathrm{S}}$ and $O^{(l)}_1$ for $l=1,2,\cdots, L$ are drawn from a Gaussian distribution with the following likelihood: 
\begin{equation*}
\label{joint:Gaussian:den}
\phi(\widehat{\beta}_{\mathrm{S}}; J\beta_{\mathbb{M}_{\mathrm{\mathrm{S}}}, \mathrm{\mathrm{S}}} + k, \Sigma)\cdot \prod_{l=1}^L \phi(O^{(l)}_1; A^{(l)} \widehat{\beta}_{\mathrm{S}} + b^{(l)}, \bar{\Sigma}^{(l)}) 
\end{equation*}
For a convex and compact set $\mathrm{R}$, $\log\mathbb{P}\left[\begin{pmatrix}\widehat{\beta}_{\mathrm{S}}^\intercal & (O^{(1)}_1)^\intercal &  \cdots & (O^{(L)}_1)^\intercal  \end{pmatrix}^\intercal \in \mathrm{R}\right]$ is bounded from above by
\begin{equation*}
\begin{aligned}
&- \displaystyle\inf_{(\widetilde{\beta}_{\mathrm{S}}, o^{(1)}_1,\cdots, o^{(L)}_1)\in \mathrm{R}}\;\Big\{\dfrac{1}{2} (\widetilde{\beta}_{\mathrm{S}}-  J\beta_{\mathbb{M}_{\mathrm{\mathrm{S}}}, \mathrm{\mathrm{S}}} - k)^{\intercal}  \Sigma^{-1}(\widetilde{\beta}_{\mathrm{S}}- J\beta_{\mathbb{M}_{\mathrm{\mathrm{S}}}, \mathrm{\mathrm{S}}} - k)\\
&\;\;\;\;\;\;\;\;\;\;\;\;\;\;\;\;\;\;\;\;\;\;+ \displaystyle\sum_{l=1}^L \frac{1}{2} (o^{(l)}_1  -A^{(l)}\widetilde{\beta}_{\mathrm{S}} -b^{(l)})^{\intercal} (\bar{\Sigma}^{(l)})^{-1}(o^{(l)}_1-A^{(l)}\widetilde{\beta}_{\mathrm{S}} -b^{(l)})\Big\}.
\end{aligned}
\end{equation*} 
\end{proposition}

The proof of the proposition closely follows the steps in the proof for Proposition 3.1; 
we thus omit the proof here.

\begin{theorem}
\label{separability:multi}
Consider solving
\vspace{-2mm}
 $$o^{*(l)}_1( \widehat{\beta}_{\mathrm{S}} ) = \; \underset{{o_1^{(l)}}}{\text{argmin}}\;\dfrac{1}{2}(o_1^{(l)}- A^{(l)}\widehat{\beta}_{\mathrm{S}}  -b^{(l)})^{\intercal}(\bar{\Sigma}^{(l)})^{-1}(o_1^{(l)}- A^{(l)}\widehat{\beta}_{\mathrm{S}}  -b^{(l)})+\mathcal{B}_{U^{(l)}; v^{(l)}}(o^{(l)}_1).
 \vspace{-2mm}
 $$
 for $l\in\{1,2,\cdots, L\}$.
Then, the maximizer of the approximate likelihood in \eqref{barrier:approx:multi} is
\begin{equation*}
\begin{aligned}
\widehat{\beta}^{\;\text{mle}}_{\mathbb{M}_{\mathrm{S}},\mathrm{S}}  &= J^{-1} \widehat{\beta}_{\mathrm{S}}- J^{-1}k +  \Sigma_{\mathbb{M}_{\mathrm{S}}, \mathrm{S}} \sum_{l=1}^{L} A^{(l)T} (\bar{\Sigma}^{(l)})^{-1} (A^{(l)} \widehat{\beta}_{\mathrm{S}} + b^{(l)} -o^{*(l)}_1(\widehat{\beta}_{\mathrm{S}})).
\end{aligned}
\end{equation*}
\end{theorem}

\begin{theorem}
\label{Fisher:info:multi}
Consider solving the $L$ optimization problems in Theorem \ref{separability:multi}.
 Then, the observed Fisher information matrix for the approximate likelihood \eqref{barrier:approx:multi} is
\vspace{-4.5mm}
\begin{align*}
&\Sigma_{\mathbb{M}_{\mathrm{S}}, \mathrm{S}}^{-1} \Big(\Sigma^{-1} + \displaystyle\Big\{\sum_{l=1}^{L}(A^{(l)})^{\intercal} (\bar{\Sigma}^{(l)})^{-1} A^{(l)}   \\
&\;\;\;\;\;\;\;\;\;\;\;\;\;\;\;\;\;-(A^{(l)})^{\intercal} (\bar{\Sigma}^{(l)})^{-1} \Big((\bar{\Sigma}^{(l)})^{-1}+ \grad^2 \mathcal{B}_{U^{(l)}; v^{(l)}}(o^{*(l)}_1(\widehat{\beta}_{\mathrm{S}})\Big)^{-1} (\bar{\Sigma}^{(l)})^{-1}A^{(l)}\Big\}\Big)^{-1} \Sigma_{\mathbb{M}_{\mathrm{S}}, \mathrm{S}}^{-1}.
\end{align*}
\end{theorem}

We provide our steps for maximum likelihood inference in Algorithm \ref{alg:conf:int:multi}. 
Our approximate selective MLE is a linear combination of the solutions from $L$ separable optimization problems, which we state in Step $\text{\textbf{($\mathbf{O^{(l)}}$)}}$ of the algorithm.
In Section 5 of the paper, we apply the algorithm to the analysis of a cancer gene expression data for eliminating selection bias from multiple queries.

\begin{algorithm}[H]
\caption{ALGORITHM 2: Approximate maximum likelihood inference post multiple convex queries}
 \label{alg:conf:int:multi}
\begin{spacing}{1.9}
\begin{algorithmic} 
\REQUIRE $L$ Queries, $\omega^{(l)} \stackrel{\text{i.i.d.}}{\sim}N(0, \Sigma_{\mathbb{W}}),$ $l=1,2, \cdots, L$
\ENSURE $\widehat{\mathrm{S}}= \mathrm{S}(\mathrm{S}^{(1)}, \cdots, \mathrm{S}^{(L)})$
\STATE {\text{Implied parameters \textbf{(P)}}}: Compute matrices: $\bar{\Sigma}^{(l)}$, $A^{(l)}$, $b^{(l)}$ for $l$ in $\{1,2, \cdots, L\}$, $\Sigma$,$J$, $k$
\FORALL {$l$ in $\{1,2, \cdots, L\}$}
\STATE {\text{Optimization \textbf{$\mathbf{(O^{(l)})}$}}}: \  $o^{*(l)}_1( \widehat{\beta}_{\mathrm{S}} ) = \underset{o_1^{(l)}}{\text{argmin}}\;\;\Big\{\dfrac{1}{2}(o_1^{(l)}- A^{(l)}\widehat{\beta}_{\mathrm{S}}  -b^{(l)})(\bar{\Sigma}^{(l)})^{-1}(o- A^{(l)}\widehat{\beta}_{\mathrm{S}}  -b^{(l)})$ \\ $\hspace{78mm}+\mathcal{B}_{U^{(l)}; v^{(l)}}(o_1^{(l)})\Big\}.$
\ENDFOR
\STATE{\text{Selective MLE \textbf{(S-MLE)}}}: $\widehat{\beta}^{\;\text{mle}}_{\mathbb{M}_{\mathrm{S}},\mathrm{S}}  = J^{-1} \widehat{\beta}_{\mathrm{S}}- J^{-1}k $ \\ \hspace{55mm} $  +\Sigma_{\mathbb{M}_{\mathrm{S}}, \mathrm{S}} \sum_{l=1}^{L} A^{(l)T} (\bar{\Sigma}^{(l)})^{-1} (A^{(l)} \widehat{\beta}_{\mathrm{S}} + b^{(l)} -o^{*(l)}_1(\widehat{\beta}_{\mathrm{S}}))$
\STATE{\text{Inverse info \textbf{(FI)}}}: ${I^{-1}(\widehat{\beta}^{\;\text{mle}}_{\mathbb{M}_{\mathrm{S}},\mathrm{S}} )=\Sigma_{\mathbb{M}_{\mathrm{S}}, \mathrm{S}}  \Big(\Sigma^{-1} + \sum_{l=1}^L \Big\{A^{(l) T} (\bar{\Sigma}^{(l)})^{-1} A^{(l)}}$\\
$\hspace{30mm}{-  A^{(l) T} (\bar{\Sigma}^{(l)})^{-1} \Big\{(\bar{\Sigma}^{(l)})^{-1}+ \grad^2 \mathcal{B}_{U^{(l)}; v^{(l)}}(o_1^{*(l)}(\widehat{\beta}_{\mathrm{S}}))\Big\}^{-1} (\bar{\Sigma}^{(l)})^{-1}A^{(l)}\Big\}\Big) \Sigma_{\mathbb{M}_{\mathrm{S}}, \mathrm{S}} }$
\STATE{\text{MLE-based inference}}:
\FORALL {$j$ in selected set $\mathrm{E}$}
\STATE{\bf{(\text{p-value for }$\beta_{j; \mathbb{M}_{\mathrm{S}}, \mathrm{S}}$)} }: $\Scale[0.95]{2\min\left(\bar{\Phi}\left(\widehat{\beta}_{j;\mathbb{M}_{\mathrm{S}},\mathrm{S}}^{\;\text{\;mle}}/\sqrt{I_{j,j}^{-1}(\widehat{\beta}^{\;\text{mle}}_{\mathbb{M}_{\mathrm{S}},\mathrm{S}} )}\right), \Phi\left(\widehat{\beta}_{j;\mathbb{M}_{\mathrm{S}}, \mathrm{S}}^{\;\text{\;mle}}/\sqrt{I_{j,j}^{-1}(\widehat{\beta}^{\;\text{mle}}_{\mathbb{M}_{\mathrm{S}},\mathrm{S}} )}\right) \right)}$
\STATE{\bf{(\text{interval for} $\beta_{j; \mathbb{M}_{\mathrm{S}}, \mathrm{S}}$)} }: $\Scale[0.95]{\left(\widehat{\beta}_{j;\mathbb{M}_{\mathrm{S}}, \mathrm{S}}^{\;\text{\;mle}}- z_{1-q/2}\cdot \sqrt{I_{j,j}^{-1}(\widehat{\beta}^{\;\text{mle}}_{\mathbb{M}_{\mathrm{S}},\mathrm{S}} )}, \widehat{\beta}_{j;\mathbb{M}_{\mathrm{S}},\mathrm{S}}^{\;\text{\;mle}}+ z_{1-q/2}\cdot \sqrt{I_{j,j}^{-1}(\widehat{\beta}^{\;\text{mle}}_{\mathbb{M}_{\mathrm{S}},\mathrm{S}} )}\right)}$
\ENDFOR
\end{algorithmic}
\end{spacing}
\end{algorithm}

\section{Simulation Experiments}
\label{simulation}

We explore the potential of our method for a wide range of signal-to-noise ratio (SNR) values in a linear regression setting. 
Following closely the setup in \cite{hastie2017extended}, in every round of experiment,  we simulate each $\mathbb{R}^p$-valued row of the design matrix $X$ from $N( 0, \Sigma(\rho))$ where the $(i,j)$-th entry of $\Sigma(\rho)= \rho^{|i-j|}$. We then draw the response as: $Y\lvert X \sim N(X\beta, \sigma^2 I)$. 
Fixing $n=300$, $p=100$, $\rho=0.35$, we consider a linearly-varying coefficient vector with $s=6$ non-zero equally spaced components that have magnitudes: $-10$,$-6$,$-2$,$2$,$6$,$10$.
We vary the noise level $\sigma^2$ to match the SNR value: $\text{SNR}=\sigma^{-2}\cdot(\beta^{\intercal}\Sigma \beta)$, and vary SNR in the set $\{0.15, 0.21, 0.26, 0.31,  0.42, 0.71, 1.22, 2.07, 3.52\}$.

Dividing our experiments into two regimes, namely randomized and non-randomized, we run a canonical LASSO query (without randomization)
\begin{equation}
\label{canonical:lasso:query}
\underset{o \in \real^p}{\text{minimize}}\;\dfrac{1}{2}\|y - Xo\|_2^2 +\lambda \|o\|_1,
\end{equation}
and a randomized LASSO query in \eqref{first:eg} with $\omega\sim N(0, \eta^2 I_p)$ and $\epsilon=n^{-1/2}$.
The randomization variance $\eta^2$ is chosen so that $(\hat{\sigma})^{-2}\eta^2= 0.50$, using the estimated noise level in the data: $\hat{\sigma}^2 =(n-p)^{-1} \|(I- X(X^T X)^{-1} X^T) y\|^2$. 
Based on an asymptotic equivalence between data splitting and a Gaussian randomization scheme \citep[see Proposition 4.1][]{selective_bayesian}, our choice of randomization variance roughly matches the amount of information used up in selection when two-thirds of the samples are allocated for the LASSO.
For each query, we carry out three common schemes to choose $\lambda$. 
Our first choice is a theoretical value proposed in \cite{negahban2009unified} and is given by $\lambda_{\text{theory}}=  \mathbb{E}[\|X^T \Psi\|_{\infty}]$ where $\Psi \sim \mathbb{N}(0, \hat\sigma^2 I)$. 
Our second and third choices are obtained from cross-validation. 
Denoted by $\lambda_{\text{cv.min}}$ and $\lambda_{\text{cv.1se}}$, the tuning parameters are associated with the lowest cross-validated error and error within 1 standard error of the best model, respectively. 

\subsection{Methods and metrics} 

In our experiments, we illustrate maximum likelihood inference for two sets of parameter vectors after selection: (i) the partial regression coefficients in the selected model, obtained by letting $\mathcal{H}(\mathrm{S})= \mathrm{E}$ and $\mathcal{F}_{\mathrm{S}} = (X_{\mathrm{E}}^\intercal X_{\mathrm{E}})^{-1}X_{\mathrm{E}}^\intercal$; (ii) the selected set of parameters in the full model, obtained by letting $\mathcal{H}(\mathrm{S})= \{1,2,\cdots, p\}$ and $\mathcal{F}_{\mathrm{S}} =\mathcal{L}_{\mathrm{E}} (X^\intercal X)^{-1}X^\intercal$, where $\mathcal{L}_{\mathrm{E}} \in \real^{|\mathrm{E}|\times p}$ is a matrix of all zeros except for the indices 
\[\mathcal{L}_{\mathrm{E}}(k,j_k) = 
      1   \ \ \ \text{for }\   k\in \{1,\cdots, |\mathrm{E}|\},\  \mathrm{E}= \{j_1,\cdots, j_{|\mathrm{E}|}\}.
\]
We call the former parameter vector ``Partial" and  the latter ``Full" in our depictions. 

In the linear regression setting described above, we compare the relative performance of our proposed method with ``Lee et al.", ``Liu et al.", ``Split", and ``Naive". 
Our method follows Algorithm \ref{alg:conf:int} with the barrier penalty 
\[\mathcal{B}_{U; v}(o_1) = \textstyle\sum_{j} \log\left(1 + \dfrac{1}{v_j- U^{\intercal}_j o_1}\right),\]
where $U_j$ is the $j^{\text{th}}$ row of $U$ and $v_j$ is the $j^{\text{th}}$ component of $v$.
First, we report the average coverage of the interval estimates produced by each strategy across all simulations.
Each simulation records the proportion of intervals that cover our target parameters in a single round of experiment. 
The nominal level of FCR aimed by the interval estimates in $10\%$.
Note, selective inference produced by  ``Liu et al." is tied only to the ``Full" parameter vector. 
The coverage for ``Naive" intervals, not adjusted for any selection, underscores the extent of selection bias for a specific value of SNR.
Next, we provide a breakdown of all methods in terms of their inferential power. 
To this end, we record the average lengths of the interval estimates and their power which we define to be the proportion of signals detected by a strategy from the ones successfully screened by the query. 
Detected variables here count the variables for which the corresponding interval estimates do not cover zero. 
Faced with two different regimes, we depict the power comparisons for the randomized and non-randomized estimates under certain best-case scenarios for each estimate.
The best-case scenarios in our experiment are led by an assessment of predictive risks for a point estimate associated with each strategy, measured through the relative risk metric:
$$\mathcal{R}(\widehat{o}^{\lambda}, \beta)= (\beta^\intercal \Sigma \beta)^{-1}\cdot \left\{{(\widehat{o}^{\lambda}- \beta)^{\intercal} \Sigma (\widehat{o}^{\lambda}- \beta)}\right\},$$ 
where $\widehat{o}^{\lambda}$ is the estimate and $\beta$ is the parameter vector.
We consider the LASSO solution as a natural point estimate for the parameter vector $\beta$ in the non-randomized regime and associate the LASSO estimate with both strategies ``Lee et al." and ``Liu et al.".
For ``Split", we consider the least squares estimate obtained after refitting the selected model to the remaining one-third of the data samples and append it with zeros for the indices not selected by the LASSO.
Note, the selective MLE, an immediate byproduct of Algorithm \ref{alg:conf:int}, appended with zeros for the inactive indices serves as a point estimate for our proposal.
We discuss the detailed findings of our experiments next.

\subsection{Findings and interpretation}

\begin{figure}[H]
  \centering
  \begin{tabular}{@{}c@{}}
    \includegraphics[width=0.90\linewidth,height=100pt]{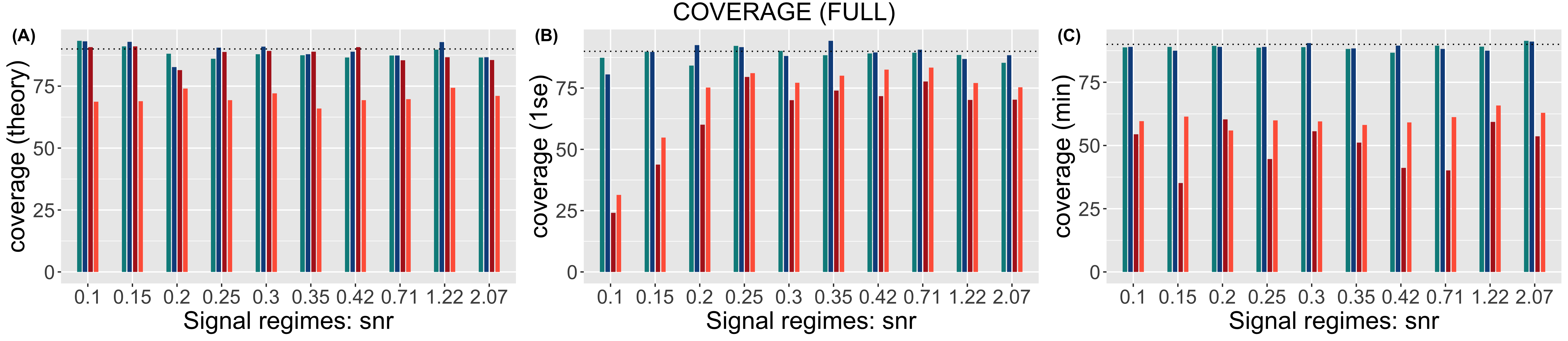} \\
  \end{tabular}
  \vspace{\floatsep}
  \begin{tabular}{@{}c@{}}
    \includegraphics[width=0.90\linewidth,height=105pt]{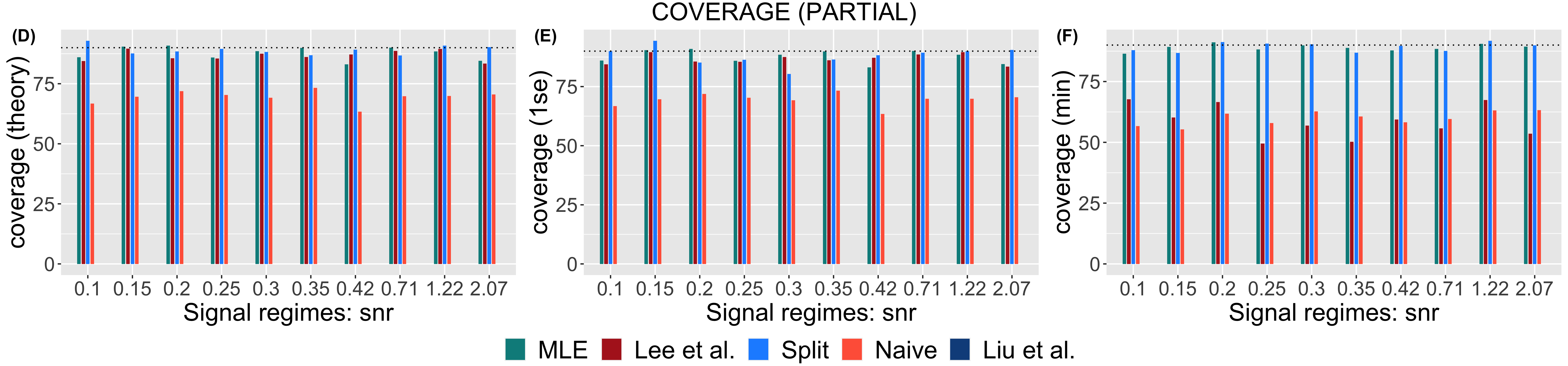} \\
  \end{tabular}
  \caption{\small{Averaged coverage of interval estimates. The nominal target for coverage is $90\%$, marked by the
  dotted horizontal line. (A), (B), (C) depict coverage for ``Full" parameters; (D), (E), (F) depict coverage for ``Partial" parameters.}}
  \label{selected:medium:coverage:n300}
\end{figure}

Supporting the validity of inference after selection, Figure \ref{selected:medium:coverage:n300} highlights the performance of the different interval estimates for the ``Full" and ``Partial" parameters. 
The three columns in the plot are associated with the three different choices of tuning parameter: $\lambda_{\text{theory}}$, $\lambda_{\text{cv.1se}}$, and $\lambda_{\text{cv.min}}$.  
We remark that none of the methods adjust for the adaptivity involved in the choice of the cross-validated tuning parameters.
Coherent with expectations, the interval estimates for all the methods approximately attain the nominal FCR level $10\%$ at $\lambda_{\text{theory}}$; ``Lee et al." fails to yield valid inference at $\lambda_{\text{cv.1se}}$ and $\lambda_{\text{cv.min}}$ since it does not account for the fact that the tuning parameters for the LASSO were chosen based on the specific data through cross-validation. 
The effectiveness of the normal approximation for the proposed method ``MLE" is largely ascribed to the soft-truncated likelihood, due to the use of randomization in the LASSO query. 
Besides, the accuracy of the large deviations-type approximation for the exact likelihood is maintained under moderate dimensions.
Interestingly, ``MLE" and ``Liu et al." recover the nominal levels even at the cross validated tuning parameter; this is seen in the second and third columns of the plot. 
Although there lacks a formal justification for this observation, heuristically, the use of randomization in the LASSO query limits the role of the data-dependent regularization. 
A similar justification possibly holds for ``Liu et al.", which achieves the same goal by constructing inference for parameters that are less affected by selection.

\begin{figure}[H]
  \centering
  \begin{tabular}{@{}c@{}}
    \includegraphics[height=165pt,width=0.88\linewidth]{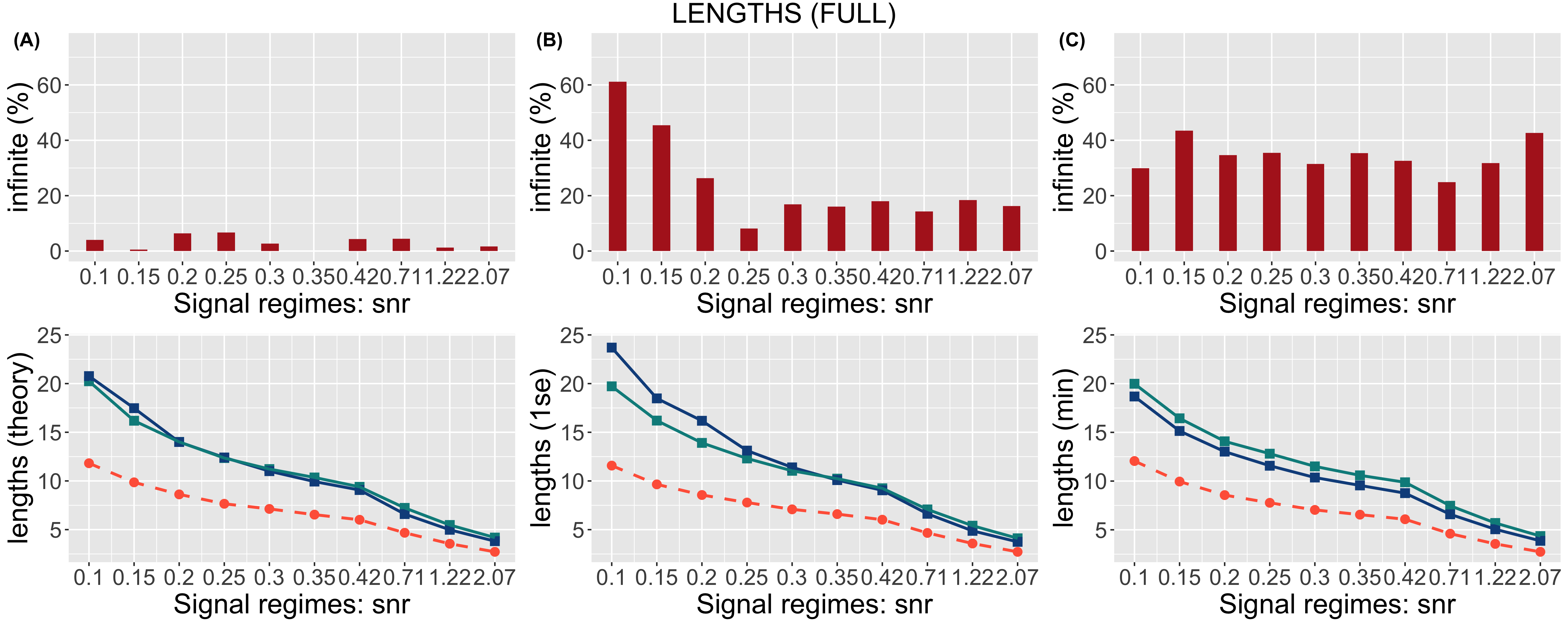} \\
  \end{tabular}
  \vspace{\floatsep}
  \begin{tabular}{@{}c@{}}
    \includegraphics[height=165pt,width=0.88\linewidth]{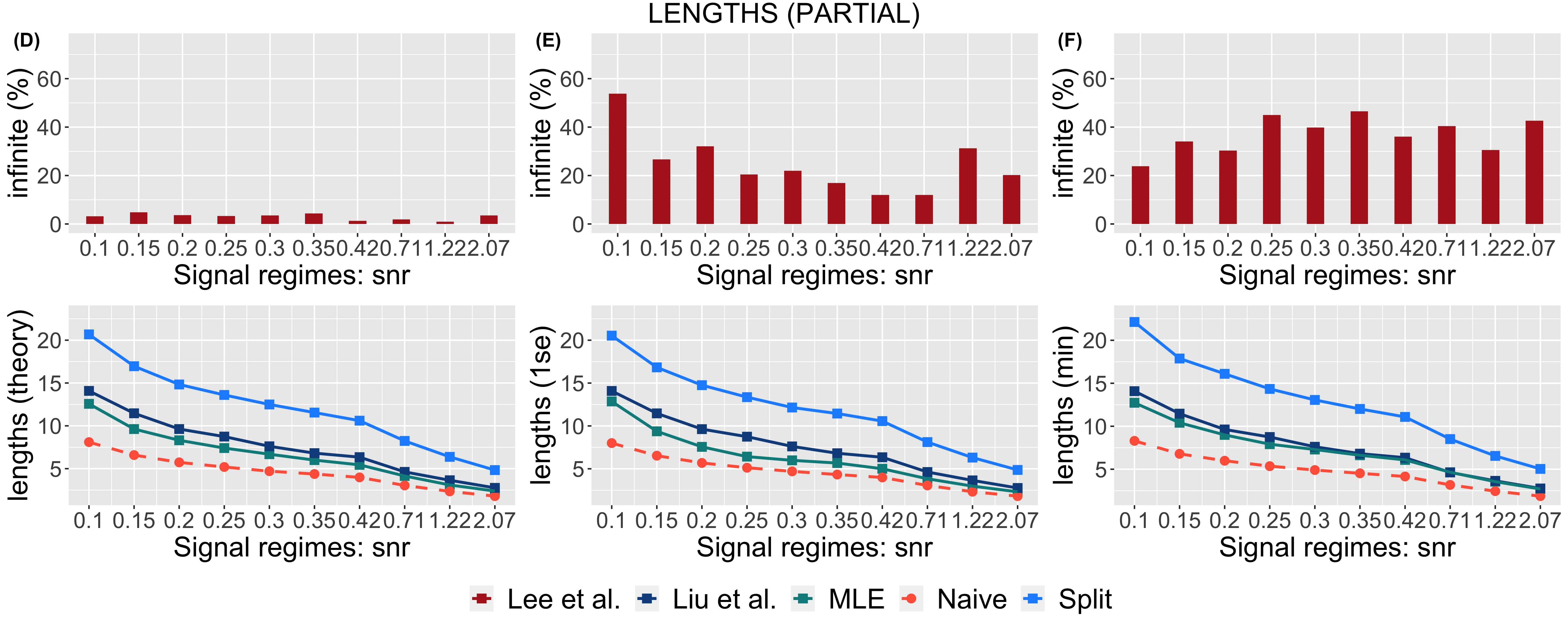} \\
  \end{tabular}
  \caption{\small{Average lengths of interval estimates. (A), (B), (C) depict the averaged lengths of intervals for ``Full" parameters; (D), (E), (F) depict the averaged lengths of intervals for ``Partial" parameters. The red bars on the top row of each panel reports the percentage of intervals by ``Lee et al." which resulted in infinite length.}}
  \label{selected:medium:length:n300}
\end{figure}

Figure \ref{selected:medium:length:n300} highlights the averaged lengths of the interval estimates produced by different methods.
The bars in red depict the percentage of intervals with infinite length for ``Lee et al.", confirming the conclusions in \cite{kivaranovic2018expected}.
The interval estimates produced by all other methods are bounded in length.
Consistent with the example presented in the introduction, the interval estimates based on ``Liu et al." and ``MLE" are comparable for the ``Full" parameters.
The estimates for the ``Partial" parameters produced by ``MLE" are shorter than those for the ``Full" parameters; this gain in power is in part due to inference in the learned model as opposed to the full model.
Assuredly, the new proposal dominates the simple ``Split" estimates which roughly use the same information in selection as the randomized query in our method.

\begin{figure}[H]
\begin{center}
\centerline{\includegraphics[width=0.95\linewidth,height=110pt]{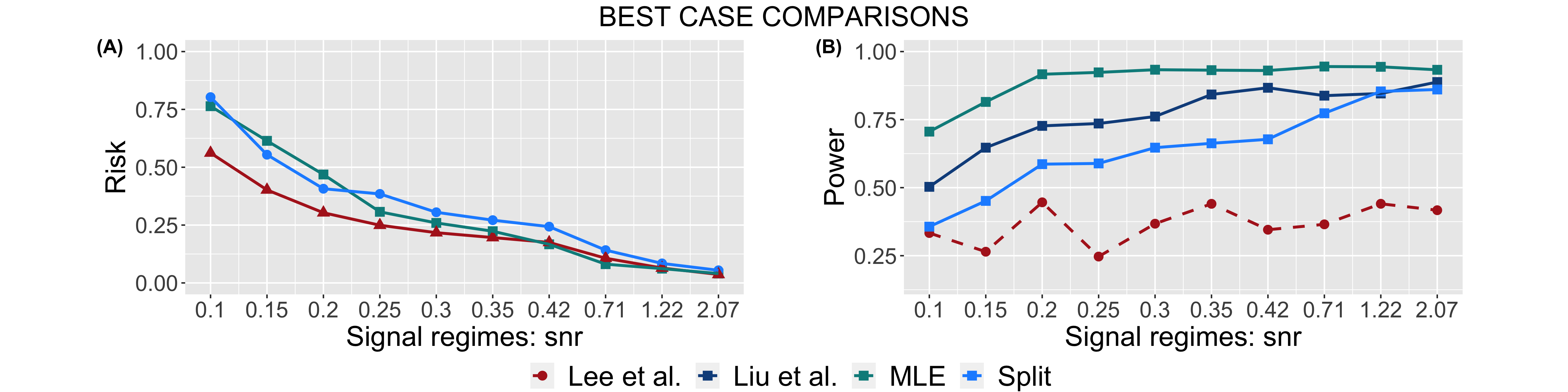}}
\end{center}
\caption{\small{Best-case Comparisons between Randomized and Non-randomized Estimates. (A) depicts the relative-risks for point estimates associated with each method; (B) depicts power of each method in detecting true associations after selection.}}
\label{best-best}
\end{figure}

Emphasized earlier, for a fair comparison of power between the randomized and non-randomized estimates, we use their relative risks to guide us to a best-case scenario within each regime.
Purely from a predictive lens, the risk assessment across the different values of SNR suggest running the randomized LASSO at $\lambda_{\text{cv.1se}}$ and the canonical LASSO at $\lambda_{\text{cv.min}}$.
Taking on direct comparisons for ``MLE" after the randomized LASSO at $\lambda_{\text{cv.1se}}$, and ``Liu et al.", ``Lee et al.", ``Split" after the usual LASSO at $\lambda_{\text{cv.min}}$, 
Figure \ref{best-best} depicts their relative risks and power under the respective best-case situations.
For the moderately high SNR values, the selective MLE proves to be a competing estimate when compared against the LASSO estimate.
We note a far superior predictive performance of the LASSO at $\lambda_{\text{cv.min}}$ than the selective MLE in the lower range of SNR values.
Our proposal, however, turns out to be a better choice for inference across the range of SNR values, outperforming the non-randomized alternatives in terms of power.

\section{Real Data Example}
\label{real:data}

We apply our method to investigate associations between gene expressions and patient survival times for Gliomas (a type of brain tumor) in the TCGA data.
With survival times ranging between $1$ to $15$ years, and some of these tumors quickly progressing to Glioblastoma, genetic associations are increasingly utilized for prognostic decisions \citep[e.g.,][]{zhang2019radio, panigrahi2020integrative}. 
In our analysis of $441$ samples, we use log-transformed survival times as our response.
As potential predictors, we choose the top $2500$ predictors with the largest sample variation from a candidate pool of $17500$ molecular measurements of gene expression values (mRNAseq). 
Before running a meaningful LASSO query, we account for the presence of strongly correlated predictors by further pruning the $2500$ predictors to a subset of $140$ predictors.
We do so by applying the hierarchical clustering scheme in \cite{bien2011hierarchical}, followed by collecting the prototype representatives for each resulting cluster of predictors.
We consider the following algorithms: (i) the LASSO; (ii) two runs of the LASSO; (iii) a marginal screening of the predictors followed by the LASSO, adding a Gaussian randomization variable $\omega\sim N(0, \eta^2 I_p)$ to the queries for our method.
Consistent with the simulations, we fix $(\hat{\sigma})^{-2}\eta^2= 1$.  
We use $\lambda_{\text{theory}}$ in Section \ref{simulation} to tune the LASSO penalty. 
We conduct a marginal screening of variables at the nominal level $q=0.20$ and let the screening threshold be $\zeta= z_{1-q/2}\cdot \sqrt{\hat{\sigma}^2 \text{diag}(X^T X) + \eta^2 1_p}$ for the randomized version of this query. 

\begin{figure}[H]
\begin{center}
\centerline{\includegraphics[height=7.3cm,width=15cm]{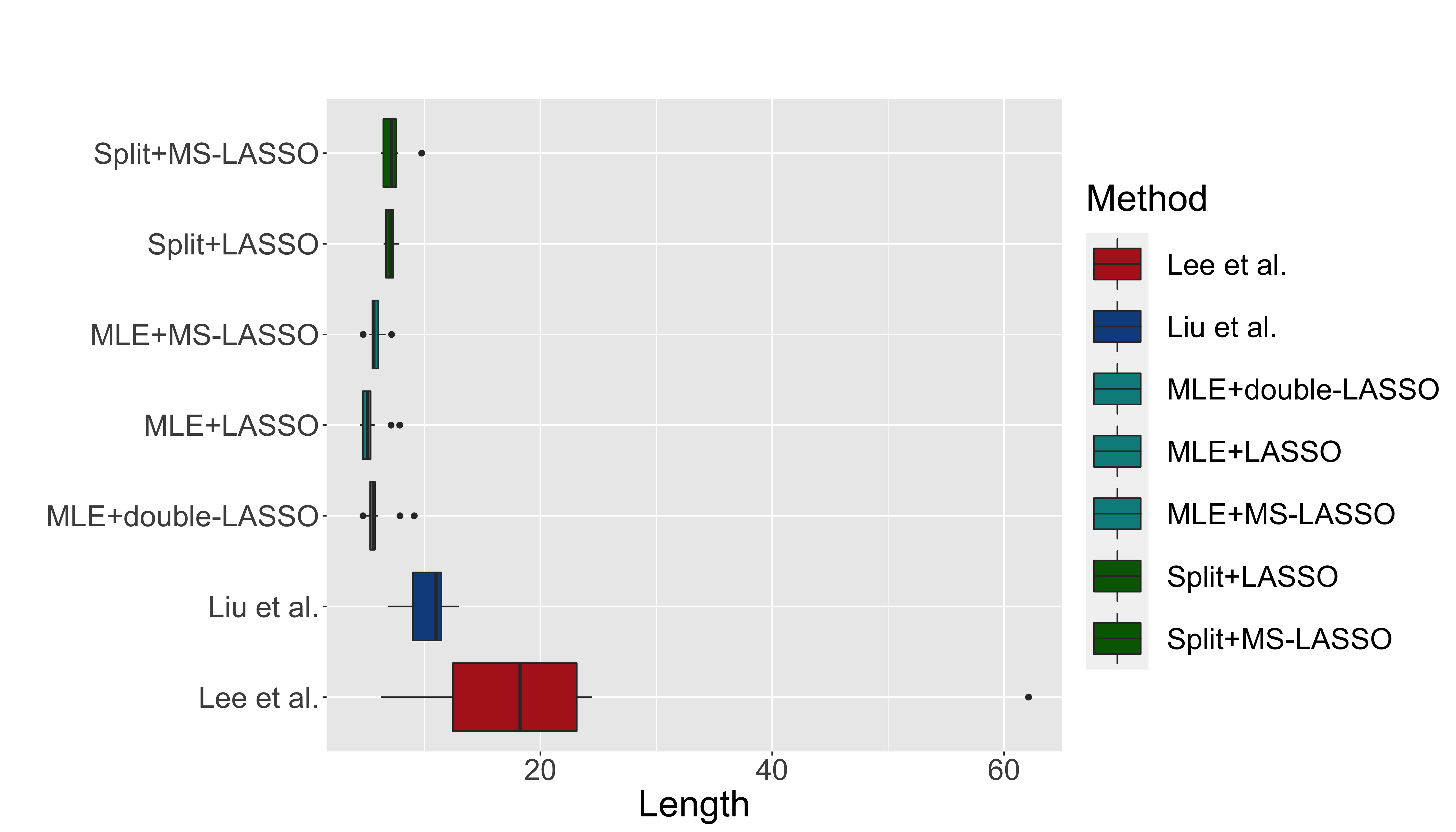}}
\end{center}
 \caption{\small{TCGA analysis. Box plots for lengths of interval estimates by all methods.}}
\label{lengths:TCGA}
\end{figure}

Figure \ref{lengths:TCGA} showcases the distribution of the lengths of interval estimates produced by ``MLE", ``Lee et al." and ``Liu et al." after solving (i). 
For inference post (ii) and (iii), we compare ``MLE $+$ double-LASSO" and ``MLE $+$ MS-LASSO" against ``Split $+$ LASSO" where half of the samples are reserved for inference. 
The conditional prescriptions in ``Lee et al." and ``Liu et al." do not directly apply to accommodate multiple queries at the time of selection. 
Corroborating our findings in the simulations, the lengths of the estimates using our proposal are way shorter than those based on ``Liu et al." and ``Lee et al.".
Observe, ``MLE $+$ double-LASSO" and ``MLE $+$ MS-LASSO" outperform ``Split $+$ LASSO" with shorter intervals, despite querying the data twice before inference. 

\section{Discussion}
\label{sec:discussion}
We investigate in the current paper a method for selective inference via maximum likelihood estimation.
Amenable to a large class of convex queries at the time of selection, we rely on an optimization problem whose solution yields us estimating equations for the MLE and the observed Fisher information matrix, the two main ingredients for the proposed method. 
The estimating equations easily generalize to multiple convex queries at the time of selection and assume a separable form across the queries.
The appeal of our method is two fold: (1) the computing costs for selective inference are reduced by orders of magnitude in comparison with MCMC sampling-based alternatives, and (2) statistical power for inference is preserved despite querying the data multiple times through randomized queries.

Future extensions of our proposal include a development of theory to use the method beyond Gaussian data. 
Along this direction, uniform guarantees for coverage \citep[e.g.,][]{leeb2005model, leeb2006can} require closer investigation.
The framework for selective inference in the paper adjusts for queries with affine K.K.T. conditions at the solution. 
The ability of the proposal to account for learning queries that present non-affine representations for the K.K.T. conditions, e.g., the Group LASSO, remain to be explored in the future.

\section{Acknowledgements}
S.P. would like to sincerely thank and acknowledge Veera Baladandayuthapani and Yujia Pan for their inputs in the analysis of the TCGA dataset. 
S.P.  is immensely thankful to Xuming He and Liza Levina for offering valuable comments on an initial draft of the paper.
The authors thank the anonymous reviewers for their many insightful suggestions on earlier drafts of the paper.

\appendix
\section{Proofs}
\label{appendix:A}

\begin{proof}
Proposition 2.1.\ \  
Recall, $$\alpha(\beta) = \dfrac{1}{2}\beta^2 + \log\bar{\Phi}\left(\frac{(\tau-\beta)}{\sqrt{(1+\eta^2)}}\right).$$
Then, we observe $\grad^2 \alpha(\beta)$ is equal to:
\begin{equation*}
\begin{aligned}
& 1- \dfrac{(\beta-\tau)}{(1+\eta^2)^{3/2}}\cdot
\left(\bar{\Phi}\left(\frac{(\tau-\beta)}{\sqrt{(1+\eta^2)}}\right)\right)^{-1}\cdot\phi\left(\frac{(\tau-\beta)}{\sqrt{(1+\eta^2)}};0,1\right)\\
&\ \ \ \ \ \ \ \ - \dfrac{1}{(1+\eta^2)}\cdot  \left(\bar{\Phi}\left(\frac{(\tau-\beta)}{\sqrt{(1+\eta^2)}}\right)\right)^{-2}\cdot\phi^2\left(\frac{(\tau-\beta)}{\sqrt{(1+\eta^2)}};0,1\right).
 \end{aligned}
\end{equation*}
From the above display, we note 
$\grad^2 \alpha(\beta)$ bounded from below by $$\dfrac{\eta^2}{(1+\eta^2)},$$
which implies that the function $\alpha$ is strongly convex.
Let $\alpha^*$ denote the conjugate function for $\alpha:\real\to \real$. 
Based on the estimating equation for the selective MLE in Equation 5, we now have:
$$\widehat{\beta}^{\text{\;mle}}- \beta = \grad\alpha^{-1}(Y)-\beta= \grad\alpha^{*}(Y)- \grad\alpha^{*}(\grad\alpha(\beta)).$$
Noting that the conjugate of a strongly convex function with index $K$ is $L$-Lipschitz smooth, where $L=K^{-1}$, 
we have:
\begin{equation}
\label{strong:convexity:uni}
\begin{aligned}
(\widehat{\beta}^{\text{\;mle}}- \beta)^2 &= (\grad\alpha^{*}(Y)- \grad\alpha^{*}(\grad\alpha(\beta)))^2 \\
& \leq \dfrac{1}{\eta^{4}} (1+\eta^2)^2  (Y-\grad\alpha(\beta))^2 \\
&= (B)^{-1}\ (Y-\grad\alpha(\beta))^2
\end{aligned}
\end{equation}
The finite sample bound in \eqref{strong:convexity:uni} now leads to our claim after taking conditional expectations 
on both sides of the inequality.
\end{proof}

\begin{proof} Theorem 1. \ \ 
For a fixed value $\mathrm{S}$, denote the density of $\widehat{\beta}^{\perp}_{\mathrm{S}}$ by $\ell^{\perp}(\cdot)$ and the density of $\omega$ conditional on $\widehat{\beta}_{\mathrm{S}}$ and $\widehat{\beta}^{\perp}_{\mathrm{S}}$ by $\ell_{\mathbb{W}}(\cdot\; \lvert \;\widehat{\beta}_{\mathrm{S}}, \widehat{\beta}^{\perp}_{\mathrm{S}})$.
We note that the unconditional multivariate likelihood based on the joint density of $\widehat{\beta}_{\mathrm{S}}$, $\widehat{\beta}^{\perp}_{\mathrm{S}}$ and $\omega$ is proportional to:
\begin{equation}
\label{lik:preCoV:0}
\begin{aligned}
& \exp\left(-\frac{1}{2}(\widehat{\beta}_{\mathrm{S}}- \beta_{\mathbb{M}_{\mathrm{S}}, \mathrm{S}})^{\intercal} \Sigma_{\mathbb{M}_{\mathrm{S}}, \mathrm{S}}^{-1}(\widehat{\beta}_{\mathrm{S}}- \beta_{\mathbb{M}_{\mathrm{S}}, \mathrm{S}})\right)\cdot \ell^{\perp}(\widehat{\beta}^{\perp}_{\mathrm{S}}) \cdot \ell_{\mathbb{W}}(\omega\; \lvert \;\widehat{\beta}_{\mathrm{S}}, \widehat{\beta}^{\perp}_{\mathrm{S}})\\
&\propto \exp\left(-\frac{1}{2}(\widehat{\beta}_{\mathrm{S}}- \beta_{\mathbb{M}_{\mathrm{S}}, \mathrm{S}})^{\intercal} \Sigma_{\mathbb{M}_{\mathrm{S}}, \mathrm{S}}^{-1}(\widehat{\beta}_{\mathrm{S}}- \beta_{\mathbb{M}_{\mathrm{S}}, \mathrm{S}})\right)\cdot \ell^{\perp}(\widehat{\beta}^{\perp}_{\mathrm{S}}) \cdot \exp\left(-\frac{1}{2}\omega^{\intercal} \Sigma^{-1}_{\mathbb{W}}\omega\right).
\end{aligned}
\end{equation}
In obtaining \eqref{lik:preCoV:0}, we have used the mutual independence between $\widehat{\beta}_{\mathrm{S}}$, $\widehat{\beta}^{\perp}_{\mathrm{S}}$ and $\omega$.
Consider a change of variables mapping:
$
\omega \stackrel{\pi^{-1}}{\longrightarrow} o,
$
based on the invertible mapping $\pi$:
\begin{equation*}
\begin{aligned}
\omega = \begin{pmatrix} \omega^{\intercal}_{\mathrm{E}} & \omega^{\intercal}_{E^c} \end{pmatrix}^{\intercal} &= - X^{\intercal} X_{\mathrm{E}}\widehat{\beta}_{\mathrm{S}} +  \begin{bmatrix}X_{\mathrm{E}}^{\intercal} X_{\mathrm{E}} + \epsilon I \\[2pt] X_{\mathrm{E}^c}^{\intercal} X_{\mathrm{E}} \end{bmatrix} o_1 + \begin{pmatrix} \lambda z_{\mathrm{E}} \\ o_2\end{pmatrix} + \widehat{\beta}^{\perp}_{\mathrm{S}}\\
&= \pi(o_1, o_2).
\end{aligned}
\end{equation*}
Ignoring the constant Jacobian, the likelihood function based on the density of $\widehat{\beta}_{\mathrm{S}}$, $\widehat{\beta}^{\perp}_{\mathrm{S}}$, $o_1$ and $o_2$ agrees (up to proportionality constants) with:
\begin{equation*}
\begin{aligned}
& \exp\left(-\frac{1}{2}(\widehat{\beta}_{\mathrm{S}}- \beta_{\mathbb{M}_{\mathrm{S}}, \mathrm{S}})^{\intercal} \Sigma_{\mathbb{M}_{\mathrm{S}}, \mathrm{S}}^{-1}(\widehat{\beta}_{\mathrm{S}}- \beta_{\mathbb{M}_{\mathrm{S}}, \mathrm{S}})\right)\cdot \ell^{\perp}(\widehat{\beta}^{\perp}_{\mathrm{S}})\cdot  \exp\left(-\frac{1}{2}(\pi(o_1, o_2))^{\intercal} \Sigma^{-1}_{\mathbb{W}}\pi(o_1, o_2) \right).
\end{aligned}
\end{equation*} 
To complete the proof, let:
$$\mathrm{R}_0 = \{o_1 : U o_1< v\},$$
and let $$C(\beta_{\mathbb{M}_{\mathrm{S}}, \mathrm{S}}) =\int \phi(\widetilde{\beta}_{\mathrm{S}}; J\beta_{\mathbb{M}_{\mathrm{S}}, \mathrm{S}} + k, \Sigma) \cdot f(\widetilde{\beta}_{\mathrm{S}})  d\widetilde{\beta}_{\mathrm{S}}.$$
Conditioning upon $\widehat{\mathrm{S}}(Y, W) =\mathrm{S}$ and $\widehat{\beta}^{\perp}_{\mathrm{S}}(Y)= \widehat{\beta}^{\perp}_{\mathrm{S}}$, the likelihood derived from the truncated (joint) density of $\widehat{\beta}_{\mathrm{S}}$ and $o_1$ is 
equal to: 
\begin{equation*}
\begin{aligned}
& (C(\beta_{\mathbb{M}_{\mathrm{S}}, \mathrm{S}}))^{-1}\cdot \phi(\widehat{\beta}_{\mathrm{S}}; J\beta_{\mathbb{M}_{\mathrm{S}}, \mathrm{S}} + k, \Sigma)\cdot \phi(o_1; A\widehat{\beta}_{\mathrm{S}} +b, \bar{\Sigma}) \cdot 1_{\mathrm{R}_0}(o_1)\\
\end{aligned}
\end{equation*}
In the previous display, we use the fact that our conditioning event is equivalent to the $|\mathrm{E}|$-dimensional linear constraints:
$U o_1< v$. 
In the final step, we marginalize over the optimization variables $o_1$, which gives us the expression of our soft-truncated likelihood.
\end{proof}
\medskip

\begin{proof}
Proposition 3.1. \ \ 
Based on the distribution of the Gaussian variables in the Proposition, we have the following bound:
\begin{equation}
\setlength{\jot}{10pt}
\begin{aligned}
& \log\mathbb{P}\left[\begin{pmatrix}\widehat{\beta}_{\mathrm{S}}^\intercal , O_1^\intercal \end{pmatrix}^\intercal \in \mathrm{R}\right] \\
& \leq \log\mathbb{\mathrm{E}}\Big[\exp\Big(\lambda_1^{\intercal} O_1 + \lambda_2^{\intercal} \widehat{\beta}_{\mathrm{S}} -\displaystyle\inf_{(\widetilde{\beta}_{\mathrm{S}},o_1)\in \mathrm{R}} \{\lambda_1^{\intercal} o_1+ \lambda_2^{\intercal}\widetilde{\beta}_{\mathrm{S}}\}\Big) \Big] \\
&= \displaystyle\sup_{(\widetilde{\beta}_{\mathrm{S}}, o_1)\in \mathrm{R}} \Big\{-\lambda_1^{\intercal} o_1 -\lambda_2^{\intercal}\widetilde{\beta}_{\mathrm{S}} + \log \mathbb{\mathrm{E}}\left[\exp(\lambda_1^{\intercal} O_1+ \lambda_2^{\intercal} \widehat{\beta}_{\mathrm{S}})\right]\Big\}.
\end{aligned}
\label{exp:bound}
\end{equation}
The bound above holds for all $\lambda_1, \lambda_2$. 
Optimizing over $\lambda_1, \lambda_2$ in \eqref{exp:bound} leads us to observe
\begin{equation*}
\begin{aligned}
&\log\mathbb{P}\left[\begin{pmatrix}\widehat{\beta}_{\mathrm{S}}^\intercal , O_1^\intercal \end{pmatrix}^\intercal \in \mathrm{R}\right] \\
&\leq -\displaystyle\sup_{\lambda_1, \lambda_2}\displaystyle\inf_{(\widetilde{\beta}_{\mathrm{S}}, o_1)\in \mathrm{R}}\Big\{\lambda_1^{\intercal} o_1 +\lambda_2^{\intercal}\widetilde{\beta}_{\mathrm{S}}  -\log \mathbb{\mathrm{E}}\left[\exp(\lambda_1^{\intercal} O_1+ \lambda_2^{\intercal} \widehat{\beta}_{\mathrm{S}})\right]\Big\}\\
&= -\inf_{(\widetilde{\beta}_{\mathrm{S}}, o_1)\in \mathrm{R}}\sup_{\lambda_1, \lambda_2}\Big\{\lambda_1^{\intercal} o_1 +\lambda_2^{\intercal}\widetilde{\beta}_{\mathrm{S}}-  \log \mathbb{\mathrm{E}}\left[\exp(\lambda_1^{\intercal} O_1+ \lambda_2^{\intercal} \widehat{\beta}_{\mathrm{S}})\right]\Big\}\\
&= -\inf_{(\widetilde{\beta}_{\mathrm{S}}, o_1)\in \mathrm{R}} \Big\{\dfrac{1}{2} (\widetilde{\beta}_{\mathrm{S}}- J\beta_{\mathbb{M}_{\mathrm{\mathrm{S}}}, \mathrm{\mathrm{S}}} - k)^{\intercal} \Sigma^{-1}(\widetilde{\beta}_{\mathrm{S}}- J\beta_{\mathbb{M}_{\mathrm{\mathrm{S}}}, \mathrm{\mathrm{S}}} - k) \\
&\;\;\;\;\;\;\;\;\;\;\;\;\;\;\;\;\;\;\;\;+  \dfrac{1}{2} (o_1  -A\widetilde{\beta}_{\mathrm{S}} -b)^{\intercal} \bar{\Sigma}^{-1}(o_1-A\widetilde{\beta}_{\mathrm{S}} -b)\Big\}.
\end{aligned}
\end{equation*}
The penultimate step holds due to a minimax equality, because of the convexity and compactness of $\mathrm{R}$. 
The last claim follows by plugging in the log-moment generating function for the multivariate Gaussian distribution of the variables $\widehat{\beta}_{\mathrm{S}}$ and $O_1$, followed by computing its conjugate function at $\begin{pmatrix}{\widetilde{\beta}_{\mathrm{S}}}^{\;\intercal} & o_1^{\intercal}\end{pmatrix}^\intercal$.
\end{proof}
\medskip

\begin{proof}
Theorem 2. \ \ 
Let $\eta_{\mathbb{M}_{\mathrm{S}}, \mathrm{S}} = \Sigma^{-1}(J \beta_{\mathbb{M}_{\mathrm{\mathrm{S}}}, \mathrm{\mathrm{S}}} + k)$, and let $h^*(\eta_{\mathbb{M}_{\mathrm{S}}, \mathrm{S}}, 0)$ be the convex conjugate of
$$
h(\widetilde{\beta}_{\mathrm{S}}, o_1) = \frac{1}{2} (\widetilde{\beta}_{\mathrm{S}})^{\intercal} \Sigma^{-1} \widetilde{\beta}_{\mathrm{S}} + \frac{1}{2}  (o_1 - A\widetilde{\beta}_{\mathrm{S}} -b)^{\intercal} \bar{\Sigma}^{-1}(o_1 - A\widetilde{\beta}_{\mathrm{S}} -b) + \mathcal{B}_{U; v}(o_1),
$$
when evaluated at $\begin{pmatrix} (\eta_{\mathbb{M}_{\mathrm{S}}, \mathrm{S}})^{\intercal} & 0^{\intercal}_{|\mathrm{E}|} \end{pmatrix}^{\intercal}$.
Based on the expression in Equation 15, note that our approximate log-selection probability, in terms of $\eta_{\mathbb{M}_{\mathrm{S}}, \mathrm{S}}$ is equal to:
$$
-\frac{1}{2} \eta_{\mathbb{M}_{\mathrm{S}}, \mathrm{S}}^{\intercal}\Sigma\eta_{\mathbb{M}_{\mathrm{S}}, \mathrm{S}} + h^*(\eta_{\mathbb{M}_{\mathrm{S}}, \mathrm{S}}, 0_{|\mathrm{E}|}).
$$
Then, the approximate selective MLE for $\beta_{\mathbb{M}_{\mathrm{S}}, \mathrm{S}}$ is given by
$$\widehat{\beta}^{\;\text{mle}}_{\mathbb{M}_{\mathrm{S}},\mathrm{S}}  = J^{-1}\Sigma \eta_{\mathbb{M}_{\mathrm{S}}, \mathrm{S}}^* - J^{-1}k,$$
where
\begin{equation}
\label{mle:natural}
\eta_{\mathbb{M}_{\mathrm{S}}, \mathrm{S}}^* = \underset{\eta_{\mathbb{M}_{\mathrm{S}}, \mathrm{S}}}{\text{argmin}}  \; h^*(\eta_{\mathbb{M}_{\mathrm{S}}, \mathrm{S}}, 0_{|\mathrm{E}|}) -\eta_{\mathbb{M}_{\mathrm{S}}, \mathrm{S}}^{\intercal}\widehat{\beta}_{\mathrm{S}}.
\end{equation}
In the remaining proof, we derive an expression for $\eta_{\mathbb{M}_{\mathrm{S}}, \mathrm{S}}^*$.
Introducing duplicate variables $\eta'_{\mathbb{M}_{\mathrm{S}}, \mathrm{S}}, u'$, we rewrite the problem in \eqref{mle:natural} as
$$
\underset{{\eta_{\mathbb{M}_{\mathrm{S}}, \mathrm{S}},\eta'_{\mathbb{M}_{\mathrm{S}}, \mathrm{S}},u,u'}}{\text{minimize}} \; h^*(\eta'_{\mathbb{M}_{\mathrm{S}}, \mathrm{S}}, u') -\eta_{\mathbb{M}_{\mathrm{S}}, \mathrm{S}}^{\intercal}\widehat{\beta}_{\mathrm{S}} + I_{0_{|\mathrm{E}|}}(u)
$$
with the linear constraints $\eta_{\mathbb{M}_{\mathrm{S}}, \mathrm{S}}=\eta'_{\mathbb{M}_{\mathrm{S}}, \mathrm{S}}$ and $u= u'$. 
The dual of the optimization in the preceding display is equivalent to solving
\[ \underset{{\zeta, o_1}}{\text{maximize}}\displaystyle\inf_{\eta_{\mathbb{M}_{\mathrm{S}}, \mathrm{S}},\eta'_{\mathbb{M}_{\mathrm{S}}, \mathrm{S}},u,u'} \text{Lg}(\zeta, o_1; \ \eta_{\mathbb{M}_{\mathrm{S}}, \mathrm{S}},\eta'_{\mathbb{M}_{\mathrm{S}}, \mathrm{S}},u,u'),\]
where the Lagrangian is as follows:
$$
\text{Lg}(\zeta, o_1; \ \eta_{\mathbb{M}_{\mathrm{S}}, \mathrm{S}},\eta'_{\mathbb{M}_{\mathrm{S}}, \mathrm{S}},u,u') = (\eta_{\mathbb{M}_{\mathrm{S}}, \mathrm{S}}-\eta'_{\mathbb{M}_{\mathrm{S}}, \mathrm{S}})^{\intercal}\zeta + (u-u')^{\intercal} o_1 - \eta_{\mathbb{M}_{\mathrm{S}}, \mathrm{S}}^{\intercal}\widehat{\beta}_{\mathrm{S}} + h^*(\eta'_{\mathbb{M}_{\mathrm{S}}, \mathrm{S}}, u') + I_{0_{|\mathrm{E}|}}(u).
$$

Because,
\[\displaystyle\inf_{\eta_{\mathbb{M}_{\mathrm{S}}, \mathrm{S}}} \eta_{\mathbb{M}_{\mathrm{S}}, \mathrm{S}}^{\intercal}(\zeta - \widehat{\beta}_{\mathrm{S}}) = -I_{\widehat{\beta}_{\mathrm{S}}}(\zeta),\]
\[\displaystyle\inf_{u} u^{\intercal} o_1 + I_{0_{|\mathrm{E}|}}(u) = 0,\]
\[\displaystyle\inf_{(\eta'_{\mathbb{M}_{\mathrm{S}}, \mathrm{S}},u')} -(\eta_{\mathbb{M}_{\mathrm{S}}, \mathrm{S}}^{'})^{\intercal}\zeta - (u^{'})^{\intercal} o_1+ h^*(\eta'_{\mathbb{M}_{\mathrm{S}}, \mathrm{S}}, u') = -h(\zeta, o_1),\]
the dual problem further simplifies as:
$$
\underset{\zeta, o_1}{\text{minimize}} \  h(\zeta, o_1) + I_{\widehat{\beta}_{\mathrm{S}}}(\zeta) = \underset{o_1}{\text{minimize}}\; h(\widehat{\beta}_{\mathrm{S}}, o_1).
$$

\noindent Denoting the optimal variables for $u$, $\zeta$ and $o_1$ by $u^*$, $\zeta^*$ and $o^*_1(\widehat{\beta}_{\mathrm{S}})$ respectively,  
 the K.K.T. conditions of optimality yield us:
\begin{equation}
\label{KKT:optimality}
(\eta_{\mathbb{M}_{\mathrm{S}}, \mathrm{S}}^{*}, u^*) =\grad {h^{*}}^{-1} (\zeta^*, o^*_1(\widehat{\beta}_{\mathrm{S}}))= \nabla h(\zeta^*, o^*_1(\widehat{\beta}_{\mathrm{S}})) = \nabla h(\widehat{\beta}_{\mathrm{S}}, o^*_1(\widehat{\beta}_{\mathrm{S}})),
\end{equation}
where 
\[o^*_1(\widehat{\beta}_{\mathrm{S}}) = \text{argmin}_{o_1}\;  \frac{1}{2}  (o_1 - A\widehat{\beta}_{\mathrm{S}} -b)^{\intercal} \bar{\Sigma}^{-1}(o_1 - A\widehat{\beta}_{\mathrm{S}} -b) + \mathcal{B}_{U; v}(o_1).\]
We conclude that
$\eta_{\mathbb{M}_{\mathrm{S}}, \mathrm{S}}^{*}$ is obtained by looking at the first $|\mathrm{E}|$ coordinates of $\nabla h(\widehat{\beta}_{\mathrm{S}}, o^*_1(\widehat{\beta}_{\mathrm{S}}))$. This gives us:
\[ \eta_{\mathbb{M}_{\mathrm{S}}, \mathrm{S}}^* =\Sigma^{-1}\widehat{\beta}_{\mathrm{S}} + A^{\intercal} \bar{\Sigma}^{-1} (A \widehat{\beta}_{\mathrm{S}} + b -o^*_1(\widehat{\beta}_{\mathrm{S}})),\]
which then results in the estimating equation:
\[\widehat{\beta}^{\;\text{mle}}_{\mathbb{M}_{\mathrm{S}},\mathrm{S}}  =  J^{-1}\widehat{\beta}_{\mathrm{S}} - J^{-1}k+\Sigma_{\mathbb{M}_{\mathrm{S}}, \mathrm{S}}  A^\intercal \bar{\Sigma}^{-1} (A \widehat{\beta}_{\mathrm{S}} + b -o_1^*(\widehat{\beta}_{\mathrm{S}})).
\]
\end{proof}
\medskip

\begin{proof}
Theorem 3. \ \   Let $\eta_{\mathbb{M}_{\mathrm{S}}, \mathrm{S}} = \Sigma^{-1}(J \beta_{\mathbb{M}_{\mathrm{\mathrm{S}}}, \mathrm{\mathrm{S}}} + k)$.
 We simplify our approximate expression for the log-selection probability:
\begin{equation*}
\begin{aligned}
& -\displaystyle\inf_{\widetilde{\beta}_{\mathrm{S}}, o_1}\Big\{ \dfrac{1}{2} (\widetilde{\beta}_{\mathrm{S}}- \Sigma\eta_{\mathbb{M}_{\mathrm{S}}, \mathrm{S}})^{\intercal} \Sigma^{-1}(\widetilde{\beta}_{\mathrm{S}}- \Sigma\eta_{\mathbb{M}_{\mathrm{S}}, \mathrm{S}})+ \dfrac{1}{2} (o_1  -A\widetilde{\beta}_{\mathrm{S}} -b)^{\intercal} \bar{\Sigma}^{-1}(o_1-A\widetilde{\beta}_{\mathrm{S}} -b)\\
 &\;\;\;\;\;\;\;\;\;\;\;\;\;\;\;\;\;\;\;\;\;\;\;\;\;\;\;\;\;\;\;\;\;\;\;\;\;\;\;\;\;\;\;\;\;\;\;\;\;\;\;\;\;\;\;\;\;\;\;\;\;\;\;\;\;\;\;\;\;\;\;\;\;\;\;\;\;\;+ \mathcal{B}_{U; v}(o_1)\Big\}
\end{aligned}
\end{equation*}
by writing it equivalently as the following $|\mathrm{E}|$-dimensional optimization:
\begin{equation}
\label{inter:part}
 -\displaystyle\inf_{o_1}  \dfrac{1}{2} (o_1-\widetilde{A}\eta_{\mathbb{M}_{\mathrm{S}}, \mathrm{S}} - \widetilde{b})^{\intercal} \widetilde{\Sigma}^{-1} (o_1-\widetilde{A}\eta_{\mathbb{M}_{\mathrm{S}}, \mathrm{S}} - \widetilde{b}) + \mathcal{B}_{U; v}(o_1),
\end{equation}
where
$$\widetilde{\Sigma}=\bar{\Sigma}+A\Sigma A^{\intercal}, \ \widetilde{A}=A\Sigma, \ \widetilde{b} =b.$$
We obtain the above equivalence by optimizing over the unconstrained variables, $\widetilde{\beta}_{\mathrm{S}}$ in the former objective.
The equivalent formulation then leads us to observe
\begin{equation}
\label{key:rep}
\begin{aligned}
\alpha(\eta_{\mathbb{M}_{\mathrm{S}}, \mathrm{S}})&= \dfrac{1}{2}\eta_{\mathbb{M}_{\mathrm{S}}, \mathrm{S}}^{\intercal}\Sigma \eta_{\mathbb{M}_{\mathrm{S}}, \mathrm{S}} - \dfrac{1}{2}\eta_{\mathbb{M}_{\mathrm{S}}, \mathrm{S}}^{\intercal} \widetilde{A}^{\intercal} \widetilde{\Sigma}^{-1} \widetilde{A} \eta_{\mathbb{M}_{\mathrm{S}}, \mathrm{S}} -  \eta_{\mathbb{M}_{\mathrm{S}}, \mathrm{S}}^{\intercal} \widetilde{A}^{\intercal}  \widetilde{\Sigma}^{-1}   \widetilde{b} - \dfrac{1}{2}\widetilde{b}^{\intercal}   \widetilde{\Sigma}^{-1}   \widetilde{b}\\
&\;\;\;\;\;\;\;\;\;\;\; + \Lambda^*\left(\widetilde{\Sigma}^{-1} (\widetilde{A}\eta_{\mathbb{M}_{\mathrm{S}}, \mathrm{S}}+\widetilde{b})\right),
\end{aligned}
\end{equation}
where $\Lambda^*$ is the conjugate function of  
$$\Lambda(o_1) = \frac{1}{2}o_1\widetilde{\Sigma}^{-1}  o_1  + \mathcal{B}_{U; v}(o_1).$$
Further, we note
$$\grad^2\alpha(\eta_{\mathbb{M}_{\mathrm{S}}, \mathrm{S}})=\widetilde{A}^{\intercal}  \widetilde{\Sigma}^{-1} \grad^2 \Lambda^*\left(\widetilde{\Sigma}^{-1} (\widetilde{A}\eta_{\mathbb{M}_{\mathrm{S}}, \mathrm{S}}+\widetilde{b})\right)\widetilde{\Sigma}^{-1} \widetilde{A} + (\Sigma-\widetilde{A}^{\intercal} \widetilde{\Sigma}^{-1} \widetilde{A}).$$
Clearly, 
$$\widetilde{A}^{\intercal}  \widetilde{\Sigma}^{-1} \grad^2 \Lambda^*\left(\widetilde{\Sigma}^{-1} (\widetilde{A}\eta_{\mathbb{M}_{\mathrm{S}}, \mathrm{S}}+\widetilde{b})\right)\widetilde{\Sigma}^{-1} \widetilde{A} $$
 is positive semi-definite, and $\alpha$ is strongly convex, satisfying specially,
$$\grad^2\alpha(v) \geq (A^{\intercal} \bar{\Sigma}^{-1} A + \Sigma^{-1})^{-1} = (\Sigma^{-1}_{\mathbb{M}_{\mathrm{S}}, \mathrm{S}}  + P_{\mathrm{S}}^{\intercal} \Sigma_{\mathbb{W}}^{-1}P_{\mathrm{S}})^{-1} \geq \lambda_0 \cdot I_{|\mathrm{E}|} \text{ for any arbitrary } v.
$$

\noindent Lastly, using the fact that
$$\widehat{\beta}^{\;\text{mle}}_{\mathbb{M}_{\mathrm{S}},\mathrm{S}}  = J^{-1}\Sigma \eta_{\mathbb{M}_{\mathrm{S}}, \mathrm{S}}^* - J^{-1}k$$
where $\eta_{\mathbb{M}_{\mathrm{S}}, \mathrm{S}}^* = \grad\alpha^{-1}(\widehat{\beta}_{\mathrm{S}})$,
we have:
\begin{equation*}
\begin{aligned}
\| \widehat{\beta}^{\;\text{mle}}_{\mathbb{M}_{\mathrm{S}},\mathrm{S}} -\beta_{\mathbb{M}_{\mathrm{S}}, \mathrm{S}}\|_2^2 
&\leq \lambda_1^{-2} \|\eta_{\mathbb{M}_{\mathrm{S}}, \mathrm{S}}^{*}-\eta_{\mathbb{M}_{\mathrm{S}}, \mathrm{S}}\|_2^2\\
&=\lambda_1^{-2} \|\grad\alpha^{-1}(\widehat{\beta}_{\mathrm{S}})-\eta_{\mathbb{M}_{\mathrm{S}}, \mathrm{S}}\|_2^2\\
&=\lambda_1^{-2} \|\grad\alpha^{*}(\widehat{\beta}_{\mathrm{S}})-\grad\alpha^{*}(\grad\alpha(\eta_{\mathbb{M}_{\mathrm{S}}, \mathrm{S}}))\|_2^2\\
&\leq \lambda_0^{-2}\lambda_1^{-2} \ \|\widehat{\beta}_{\mathrm{S}}- \grad\alpha(\eta_{\mathbb{M}_{\mathrm{S}}, \mathrm{S}})\|^2_2\\
\end{aligned}
\end{equation*}
The ultimate inequality relies on the $L$-Lipschitz nature of $\alpha^*$ where $L\leq (\lambda_0)^{-1}$; this fact is a direct consequence of the strong convexity noted for $\alpha$.
Taking conditional expectations on both sides gives us the finite-sample bound in our claim.
\end{proof}

\medskip

\begin{proof}
Theorem 4. \ \  
Consider the notations in the proof of Theorem 2.
Recall, the expression for our approximate log-likelihood is given by:
$$\eta_{\mathbb{M}_{\mathrm{S}}, \mathrm{S}}^{\intercal}\widehat{\beta}_{\mathrm{S}} - h^*(\eta_{\mathbb{M}_{\mathrm{S}}, \mathrm{S}}, 0_{|\mathrm{E}|}).$$
The second derivative of the negative (log-) likelihood with respect to $\eta_{\mathbb{M}_{\mathrm{S}}, \mathrm{S}}$ is equal to
\[\dfrac{\partial }{\partial \eta_{\mathbb{M}_{\mathrm{S}}, \mathrm{S}}}(\grad h^{-1}(\eta_{\mathbb{M}_{\mathrm{S}}, \mathrm{S}}, 0_{|\mathrm{E}|}))= \dfrac{\partial \widehat\beta^*_{\mathrm{S}}}{\partial \eta_{\mathbb{M}_{\mathrm{S}}, \mathrm{S}}} \]
 where $\widehat\beta^*_{\mathrm{S}}$, dependent on $\eta_{\mathbb{M}_{\mathrm{S}}, \mathrm{S}}$, satisfies the following set of equations
\begin{equation}
\label{KKT:1}
(\Sigma^{-1} + A^{\intercal} \bar{\Sigma}^{-1} A)\widehat\beta^*_{\mathrm{S}} + A^{\intercal} \bar{\Sigma}^{-1} (b- o_1^*) = \eta_{\mathbb{M}_{\mathrm{S}}, \mathrm{S}};
\end{equation}
\begin{equation}
\label{KKT:2}
\bar{\Sigma}^{-1}(o_1^* - A\widehat\beta^*_{\mathrm{S}} -b)+ \grad \mathcal{B}_{U; v}(o_1^* )=0.
\end{equation}
Based on \eqref{KKT:1}, taking a derivative with respect to $\eta_{\mathbb{M}_{\mathrm{S}}, \mathrm{S}}$, we have:
$$(\Sigma^{-1} + A^{\intercal} \bar{\Sigma}^{-1} A)\dfrac{\partial \widehat\beta^*_{\mathrm{S}}}{\partial \eta_{\mathbb{M}_{\mathrm{S}}, \mathrm{S}}} -  A^{\intercal} \bar{\Sigma}^{-1}\dfrac{\partial o_1^*}{\partial \widehat\beta^*_{\mathrm{S}}} \dfrac{ \partial \widehat\beta^*_{\mathrm{S}}} {\partial \eta_{\mathbb{M}_{\mathrm{S}}, \mathrm{S}}} = I,$$
and therefore:
$$\dfrac{\partial \widehat\beta^*_{\mathrm{S}}}{\partial \eta_{\mathbb{M}_{\mathrm{S}}, \mathrm{S}}}= \left(\Sigma^{-1} + A^{\intercal} \bar{\Sigma}^{-1} A -  A^{\intercal} \bar{\Sigma}^{-1}\dfrac{\partial o_1^*}{\partial \widehat\beta^*_{\mathrm{S}}}\right)^{-1}.$$
We obtain an expression for $\dfrac{\partial o_1^*}{\partial \widehat\beta^*_{\mathrm{S}}}$ from 
 \eqref{KKT:2} after taking a derivative with respect to $\widehat\beta^*_{\mathrm{S}}$, which gives us:
$$\left(\bar{\Sigma}^{-1}+ \grad^2 \mathcal{B}_{U; v}(o_1^*)\right)\cdot \dfrac{\partial o_1^*}{\partial \widehat\beta^*_{\mathrm{S}}} = \bar{\Sigma}^{-1}A.$$
Lastly, we note $\widehat\beta^*_{\mathrm{S}}= \widehat{\beta}_{\mathrm{S}}$ when $\eta_{\mathbb{M}_{\mathrm{S}}, \mathrm{S}} = \eta_{\mathbb{M}_{\mathrm{S}}, \mathrm{S}}^*$, the optimizer defined in  \eqref{mle:natural}.
Thus, 
$\dfrac{\partial \widehat\beta^*_{\mathrm{S}}}{\partial \eta_{\mathbb{M}_{\mathrm{S}}, \mathrm{S}}}(\eta_{\mathbb{M}_{\mathrm{S}}, \mathrm{S}})\Big\lvert_{\eta_{\mathbb{M}_{\mathrm{S}}, \mathrm{S}}= \eta_{\mathbb{M}_{\mathrm{S}}, \mathrm{S}}^*}$ is equal to
$$\left(\Sigma^{-1} + A^{\intercal} \bar{\Sigma}^{-1} A -  A^{\intercal} \bar{\Sigma}^{-1} \left(\bar{\Sigma}^{-1}+ \grad^2 \mathcal{B}_{U; v}(o_1^*(\widehat{\beta}_{\mathrm{S}})))\right)^{-1} \bar{\Sigma}^{-1}A\right)^{-1},$$
where $o_1^*(\widehat{\beta}_{\mathrm{S}})$ is the solution of \eqref{KKT:2} when $\widehat\beta^*_{\mathrm{S}}= \widehat{\beta}_{\mathrm{S}}$.
Using the reparameterization
$$\beta_{\mathbb{M}_{\mathrm{S}},\mathrm{S}}  = J^{-1}\Sigma \eta_{\mathbb{M}_{\mathrm{S}}, \mathrm{S}} - J^{-1}k = \Sigma_{\mathbb{M}_{\mathrm{S}}, \mathrm{S}} \eta_{\mathbb{M}_{\mathrm{S}}, \mathrm{S}} - J^{-1}k,$$
we deduce that observed Fisher information matrix, $I(\widehat{\beta}^{\;\text{mle}}_{\mathbb{M}_{\mathrm{S}},\mathrm{S}} )$, equals
$$
\Sigma^{-1}_{\mathbb{M}_{\mathrm{S}}, \mathrm{S}} \left(\Sigma^{-1} + A^{\intercal} \bar{\Sigma}^{-1} A -  A^{\intercal} \bar{\Sigma}^{-1} \left(\bar{\Sigma}^{-1}+ \grad^2 \mathcal{B}_{U; v}(o_1^*(\widehat{\beta}_{\mathrm{S}}))\right)^{-1} \bar{\Sigma}^{-1}A\right)^{-1} \Sigma^{-1}_{\mathbb{M}_{\mathrm{S}}, \mathrm{S}}.
\vspace{-4mm}
$$
\end{proof}

\begin{proof}
Theorem 5. \ \  
Let $\eta_{\mathbb{M}_{\mathrm{S}}, \mathrm{S}} = \Sigma^{-1}(J \beta_{\mathbb{M}_{\mathrm{\mathrm{S}}}, \mathrm{\mathrm{S}}} + k)$.
Based upon the approximation in Equation 19, the MLE problem solves the optimization
\begin{equation}
\label{mle:natural:multi}
\underset{{\eta_{\mathbb{M}_{\mathrm{S}}, \mathrm{S}}}}{\text{minimize}} -\eta_{\mathbb{M}_{\mathrm{S}}, \mathrm{S}}^{\intercal}\widehat\beta_{\mathrm{S}} + h^*(\eta_{\mathbb{M}_{\mathrm{S}}, \mathrm{S}}, 0),
\end{equation}
where $h^*(\eta_{\mathbb{M}_{\mathrm{S}}, \mathrm{S}}, 0)$ is the convex conjugate of $h$ defined as 
\begin{equation*}
\begin{aligned}
h(\widetilde{\beta}_{\mathrm{S}}, o_1^{(1)}, \cdots, o_1^{(L)}) &= \frac{1}{2} \widetilde\beta_{\mathrm{S}}^{\intercal} \Sigma^{-1} \widetilde{\beta}_{\mathrm{S}} + \displaystyle\sum_{l=1}^{L}\Big\{\frac{1}{2}  (o_1^{(l)}- A^{(l)}(\bar{\Sigma}^{(l)})^{-1}\widetilde{\beta}_{\mathrm{S}} -b^{(l)})^{\intercal} \\
&\;\;\;\;\;\;\;\;\;\;\;\;\;\;\;\;\;\;\;\;\;\;\;\;\;\;\;\;\;\;\;\;\;\;(o_1^{(l)}- A^{(l)}\widetilde{\beta}_{\mathrm{S}} -b^{(l)}) + \mathcal{B}_{U^{(l)}; v^{(l)}}(o_1^{(l)})\Big\}.
\end{aligned}
\end{equation*}
Our selective MLE is then given by:
$$\widehat{\beta}^{\;\text{mle}}_{\mathbb{M}_{\mathrm{S}},\mathrm{S}}  = J^{-1}\Sigma \eta_{\mathbb{M}_{\mathrm{S}}, \mathrm{S}}^* - J^{-1}k,$$
such that $\eta_{\mathbb{M}_{\mathrm{S}}, \mathrm{S}}^*$ is the minimizer of the optimization in \eqref{mle:natural:multi}.
Our proof is complete by noting that the dual of the optimization in \eqref{mle:natural:multi} is equal to:
\begin{equation*}
\begin{aligned}
\underset{{\zeta, o_1^{(1)}, \cdots, o_1^{(L)}}}{\text{minimize}}\;\; h(\zeta, o_1^{(1)}, \cdots, o_1^{(L)}) + I_{\widehat\beta_{\mathrm{S}}}(\zeta) &=\underset{o_1^{(1)}, \cdots, o_1^{(L)}}{\text{minimize}}\;\; h(\widehat\beta_{\mathrm{S}},  o_1^{(1)}, \cdots, o_1^{(L)}).
\end{aligned}
\end{equation*}
Observe, the objective in the dual problem is separable in the optimizing variables $o_1^{(1)}, \cdots, o_1^{(L)}$.
The K.K.T. conditions of optimality indicate that
$\eta_{\mathbb{M}_{\mathrm{S}}, \mathrm{S}}^{*}$ is obtained by looking at the first $|\mathrm{E}|$ coordinates of $$\nabla h(\widehat{\beta}_{\mathrm{S}}, o^{*(1)}(\widehat{\beta}_{\mathrm{S}}), \cdots, o^{*(L)}(\widehat{\beta}_{\mathrm{S}}))$$ 
where $o^{*(l)}_1(\widehat{\beta}_{\mathrm{S}})$ for $l\in \{1,2,\cdots, L\}$ are the optimal variables for the dual problem.
This gives us the estimating equation for the approximate selective MLE.
\end{proof}

\begin{proof}
Theorem 6. \ \  
Borrowing notations from the proof of Theorem \ref{separability:multi}, the Hessian of the negative log-likelihood
\begin{equation}
\label{mle:problem:multi} 
-\eta_{\mathbb{M}_{\mathrm{S}}, \mathrm{S}}^{\intercal}\widehat\beta_{\mathrm{S}} + h^*(\eta_{\mathbb{M}_{\mathrm{S}}, \mathrm{S}}, 0)
\end{equation} 
with respect to $\eta_{\mathbb{M}_{\mathrm{S}}, \mathrm{S}}$ is equal to
\[\dfrac{\partial}{\partial \eta_{\mathbb{M}_{\mathrm{S}}, \mathrm{S}}}(\grad h^{-1}(\eta_{\mathbb{M}_{\mathrm{S}}, \mathrm{S}}, 0))= \dfrac{\partial \widehat\beta^*_{\mathrm{S}}}{\partial \eta_{\mathbb{M}_{\mathrm{S}}, \mathrm{S}}}. \]
In the above display, $\widehat\beta^*_{\mathrm{S}}$ satisfies the following system of equations
\begin{equation}
\label{KKT:1:multi}
\left(\Sigma^{-1} + \sum_{l=1}^{L}  (A^{(l)})^{\intercal} (\bar{\Sigma}^{(l)})^{-1} A^{(l)}\right)\widehat\beta^*_{\mathrm{S}} + \sum_{l=1}^{L} (A^{(l)})^{\intercal} (\bar{\Sigma}^{(l)})^{-1} (b^{(l)}- o_1^{*(l)}) = \eta_{\mathbb{M}_{\mathrm{S}}, \mathrm{S}},
\end{equation}
and 
\begin{equation}
\label{KKT:2:multi}
(\bar{\Sigma}^{(l)})^{-1}(o_1^{*(l)} - A^{(l)}\widehat\beta^*_{\mathrm{S}} -b^{(l)})+ \grad \mathcal{B}_{U^{(l)}; v^{(l)}}(o_1^{*(l)} )=0 \text{ for } l=1,2,\cdots, L.
\end{equation}
Taking derivatives of equations \eqref{KKT:1:multi} and \eqref{KKT:2:multi}, we have 
$$\dfrac{\partial \widehat\beta^*_{\mathrm{S}}}{\partial \eta_{\mathbb{M}_{\mathrm{S}}, \mathrm{S}}}= \left(\Sigma^{-1} +  \sum_{l=1}^{L}  \left\{(A^{(l)})^{\intercal} (\bar{\Sigma}^{(l)})^{-1} A^{(l)} -  (A^{(l)})^{\intercal} (\bar{\Sigma}^{(l)})^{-1} \dfrac{\partial o_1^{*(l)}}{\partial \widehat\beta^*_{\mathrm{S}}}\right\}\right)^{-1} $$
$$\left((\bar{\Sigma}^{(l)})^{-1}+ \grad^2 \mathcal{B}_{U^{(l)}; v^{(l)}}(o_1^{*(l)})\right)\cdot  \dfrac{\partial o_1^{*(l)}}{\partial \widehat\beta^*_{\mathrm{S}}} = (\bar{\Sigma}^{(l)})^{-1}A^{(l)}.$$
Observe, $\widehat\beta^*_{\mathrm{S}}= \widehat{\beta}_{\mathrm{S}}$ when $\eta_{\mathbb{M}_{\mathrm{S}}, \mathrm{S}} = \eta_{\mathbb{M}_{\mathrm{S}}, \mathrm{S}}^*$, the minimizer for the problem in  \eqref{mle:natural:multi}.
Therefore, we have
\begin{align*}
& \dfrac{\partial \widehat\beta^*_{\mathrm{S}}}{\partial \breve{\eta}_{\mathrm{S}}}(\eta_{\mathbb{M}_{\mathrm{S}}, \mathrm{S}})\Big\lvert_{\eta_{\mathbb{M}_{\mathrm{S}}, \mathrm{S}}^*} =  \Big(\Sigma^{-1} + \displaystyle\Big\{\sum_{l=1}^{L}(A^{(l)})^{\intercal} (\bar{\Sigma}^{(l)})^{-1} A^{(l)}   \\
&\;\;\;\;\;\;\;\;\;\;\;\;\;\;\;\;\;-(A^{(l)})^{\intercal} (\bar{\Sigma}^{(l)})^{-1} \Big((\bar{\Sigma}^{(l)})^{-1}+ \grad^2 \mathcal{B}_{U^{(l)}; v^{(l)}}(o^{*(l)}_1(\widehat{\beta}_{\mathrm{S}})\Big)^{-1}  (\bar{\Sigma}^{(l)})^{-1}A^{(l)}\Big\}\Big)^{-1},
\end{align*}
where $o^{*(l)}_1(\widehat{\beta}_{\mathrm{S}})$ satisfies \eqref{KKT:2:multi} when $\widehat\beta^*_{\mathrm{S}}= \widehat{\beta}_{\mathrm{S}}$.
Finally, using the reparameterization
$$\beta_{\mathbb{M}_{\mathrm{S}},\mathrm{S}}  = J^{-1}\Sigma \eta_{\mathbb{M}_{\mathrm{S}}, \mathrm{S}} - J^{-1}k = \Sigma_{\mathbb{M}_{\mathrm{S}}, \mathrm{S}} \eta_{\mathbb{M}_{\mathrm{S}}, \mathrm{S}} - J^{-1}k,$$
we obtain our expression for the observed Fisher information matrix.
\end{proof}

\section{Soft-truncated likelihood: examples}
\label{appendix:B}

In this section, we illustrate how our method applies to: (i) variable screening based on marginal correlations \citep{lee2014exact}; (ii) variable selection via SLOPE \citep{bogdan2015slope}.

\begin{Example}
\label{MS:example:KKT}\ \ 
\rm{
In this example, we consider a randomized variable screening algorithm based on marginal correlations. 
Fix a level $q\in (0,1)$ and consider drawing a randomization variable $\omega\sim N(0_p, \eta^2 I_p)$.
Let
$$\zeta= z_{1-q/2}\cdot \sqrt{{\sigma}^2\cdot \text{diag}(X^{\intercal} X) + \eta^2\cdot 1_p}.$$
Suppose, we solve: 
 \begin{equation}
\label{marginal:screening}
\underset{o}{\text{minimize}} \; \frac{1}{2} \|o - X^{\intercal} y\|_2^2  +  \mathcal{P}^{\zeta}(o)-\omega^{\intercal} o,
\end{equation}
where $\mathcal{P}^{\zeta}(o) = \displaystyle\sum_j \chi_{\{|o_j| \leq \zeta_j\}}(o_j)$, and $\chi_R(o)$ takes the value $0$ if $o\in R$ and is infinity otherwise. 
Let the solution of \eqref{marginal:screening} be given by: $o^{(\zeta)}$.
Then, let $\mathrm{E}$ represent the indices of the selected variables, i.e., $\mathrm{E}= \{j: |o^{(\zeta)}_j|= \zeta_j\}$, and let $z_E$ be the signs of the vector: $X_E^{\intercal} y +\omega_E$. 
Let $o_1$ be the subgradient (subvector) of the penalty $\mathcal{P}^{\zeta}$ at the indices in $\mathrm{E}$, and let $o_2=X_{\mathrm{E}^c}^{\intercal} y +\omega_{\mathrm{E}^c}$.

Suppose, we condition on the observed value of $\mathrm{E}$ along with the additional conditioning on $z_E$ and $o_2$.
That is, $\mathrm{S}$ is the observed active set, alongside the sign vector $\text{sign}(X_E^{\intercal} y +\omega_E)$ and the observed vector $X_{\mathrm{E}^c}^{\intercal} y +\omega_{\mathrm{E}^c}$.
Fixing $\mathcal{H}(\mathrm{S})= \mathrm{E}$, $\mathcal{F}_{\mathrm{S}}= (X_{\mathrm{E}}^{\intercal} X_{\mathrm{E}})^{-1}X_{\mathrm{E}}^{\intercal} \in\mathbb{R}^{|\mathrm{E}| \times n}$, we work with
$\widehat{\beta}_{\mathrm{S}}$ and $\widehat{\beta}^{\perp}_{\mathrm{S}}$ defined within Section 3 for the randomized LASSO query.
Fix the following matrices
\begin{equation*} 
\label{KKT:marginal:screening}
P_{\mathrm{S}} = - \begin{bmatrix} X_E & X_{\mathrm{E}^c} \end{bmatrix}^{\intercal} X_E, \; Q_{\mathrm{S}} = \begin{bmatrix} I_{|\mathrm{E}|\times |\mathrm{E}|} \\  0_{p-|\mathrm{E}|\times |\mathrm{E}|} \end{bmatrix}, \;r_{\mathrm{S}}(\widehat{\beta}^{\perp}_{\mathrm{S}}, o_2) 
=\begin{pmatrix} \text{diag}(z_{\mathrm{E}}) \zeta_{\mathrm{E}} \\ o_2\end{pmatrix} + \widehat{\beta}^{\perp}_{\mathrm{S}}.
\end{equation*}
Then, the K.K.T mapping associated with the solution of \eqref{marginal:screening} is
$$\omega = \begin{pmatrix} \omega^{\intercal}_{\mathrm{E}} & \omega^{\intercal}_{E^c} \end{pmatrix}^{\intercal} =P_{\mathrm{S}}  \widehat{\beta}_{\mathrm{S}} + Q_{\mathrm{S}} o_1  + r_{\mathrm{S}}(\widehat{\beta}^{\perp}_{\mathrm{S}}, o_2).$$
Our event of selection is clearly equivalent to: $U o_1< v$, for the fixed matrices $U= -\text{diag}(z_{\mathrm{E}})$, $v= 0_{|\mathrm{E}|}$.
Now, we can follow the exact same analysis as prescribed for the randomized LASSO query.
}
\end{Example}

\begin{Example}
\label{SLOPE:example:KKT}\ \ 
\rm{
We consider a randomized version of the SLOPE \citep{bogdan2015slope}, given by:
\begin{equation}
\label{SLOPE}
\underset{o}{\text{minimize}} \; \frac{1}{2} \|y - X o\|_2^2 + \displaystyle\sum_{j=1}^{p} \lambda_j |o|_{(j)} -\omega^{\intercal} o.
\end{equation}
The randomization instance $\omega$ is drawn from a Gaussian distribution as our previous examples. 
The SLOPE penalty is an ordered version of the $\ell_1$ penalty. 
Assuming the $p$ tuning parameters are distinct, we let $\mathrm{E}$ be the indices of the selected variables with the sign vector $z_{\mathrm{E}}$ based on the solution of \eqref{SLOPE} . 
Because, the SLOPE penalty generates ties in the absolute values of (the nonzero components of) the solution vector, we let
 $|\breve{\mathrm{E}}|$ be the number of distinct nonzero components for the solution to \eqref{SLOPE}. 
 Consider the vector containing the magnitudes of these distinct nonzero components, $O_1\in \real^{|\breve{\mathrm{E}}|}$; that is,
$$O_1^{(1)} > O_1^{(2)} >  \cdots > O_1^{(|\breve{\mathrm{E}}|)}.
$$
Suppose, each distinct solution $O_1^{(k)}$ is associated with the following cluster of variables: $\mathcal{A}_k$.
Denote $\bar{X}_k= \sum_{k\in \mathcal{A}_k} z_k X_k$, where $z_k$ is the sign of the estimated solution for the $k^{\text{th}}$ variable.
Let $O_2$ stand for the inactive components of the subgradient for the SLOPE penalty.
We denote the observed instances for $O_1$ and $O_2$ by $o_1$ and $o_2$, respectively.

In this example, we define our conditioning event to be the set of realizations that lead us to observing the active set of variables $\mathrm{E}$, along with the signs $z_{\mathrm{E}}$ and the inactive subgradient $o_2$.
Working with the specific choices: $\mathcal{H}(\mathrm{S})= \mathrm{E}$ and $\mathcal{F}_{\mathrm{S}}= (X_{\mathrm{E}}^{\intercal} X_{\mathrm{E}})^{-1}X_{\mathrm{E}}^{\intercal}$, define $\widehat{\beta}_{\mathrm{S}}$ and $\widehat{\beta}^{\perp}_{\mathrm{S}}$
as we have done in the previous example. 
We recognize that the K.K.T. mapping at the solution is given by:
\begin{align*}
\omega = \begin{pmatrix} \omega^{\intercal}_{\mathrm{E}} & \omega^{\intercal}_{E^c} \end{pmatrix}^{\intercal}  &= - X^{\intercal} X_E \widehat{\beta}_{\mathrm{S}} + X^{\intercal}\begin{bmatrix} \bar{X}_1 & \cdots & \bar{{X}}_{|\breve{\mathrm{E}}|} \end{bmatrix}  o_1 + \widehat{\beta}^{\perp}_{\mathrm{S}} \\
&\;\;\;\;\;+ \dfrac{\partial}{\partial o}\left(\sum_{j=1}^p \lambda_j |o|_{(j)}\right)\Big \lvert_{(o^{\intercal}_1, 0_{p-|\mathrm{E}|}^{\intercal})^{\intercal}}.
\label{KKT:SLOPE}
\end{align*}
Let $V\in \real^{|\breve{\mathrm{E}}|-1 \times |\breve{\mathrm{E}}|}$ be a matrix of all zeros, except the indices
$$V[i,i] = -1, \; V[i,i+1] = 1, \text{ for } i=1,2,\cdots, |\breve{\mathrm{E}}|-1.$$
In particular, our selection event is equivalent to the following linear constraints on $o_1$: $U o_1< v$ for the fixed matrices $U= \text{diag}(-I_{|\breve{\mathrm{E}}|}, V)$, $v= 0_{2|\breve{\mathrm{E}}|-1}$.
We are now ready to apply Theorem 2 and  4 in order to obtain the estimating equations for the approximate selective MLE and observed Fisher information matrix, respectively.
}
\end{Example}

\section{Asymptotic Properties of Selective MLE}
\label{appendix:C}

We first turn to the univariate example in Section 2. 
An application of Proposition 2.1 results in a global consistency guarantee for the selective MLE that we formalize below. 
We consider the selection rule
 \begin{equation}
 \label{sel:rule:asymp:uni}
 \sqrt{n}\bar{Y}_n + W >\tau,
 \end{equation}
 such that $\sqrt{n} \bar{Y}_n \sim N(\sqrt{n}\beta_n, 1)$ and $W\sim N(0, \eta^2)$ is our randomization variable, independent of $\bar{Y}_n$. 
 We denote the corresponding selective MLE by $\widehat{\beta}_n^{\text{\;mle}}$.

\begin{proposition}
 \label{uni:consistency:mle}
Fix $\delta>0$.
Then, 
$$\mathbb{P}\left[|\widehat{\beta}_n^{\text{\;mle}}- \beta_n|>\delta  \; \lvert \; \sqrt{n}\bar{Y}_n+W>\tau \right] \to 0$$
as $n\to\infty$.
\end{proposition}

\begin{proof}
Using the bound for the mean squared error of the selective MLE in Proposition 2.1,
we have
\begin{equation*}
\label{mse:bound:consistency}
\begin{aligned}
&\mathbb{\mathrm{E}}\left[n (\widehat{\beta}_n^{\text{\;mle}}- \beta_n)^2 \; \lvert \; \sqrt{n}\bar{Y}_n+W>\tau \right] \\
&\leq  (B)^{-1}\;\text{Var}(\sqrt{n} \bar{Y}_n\; \lvert \; \sqrt{n}\bar{Y}_n+W>\tau)\\
&\leq (B)^{-1}\;\text{Var}(\sqrt{n} \bar{Y}_n)= B^{-1}.
\end{aligned}
\end{equation*}
The inequality on the last display uses a reduction in variance when the variable is truncated to a convex region \citep{kanter1977reduction}.
The proof of consistency is now immediate by applying the Chebyshev's inequality.
\end{proof}

We devote the remaining section to study the asymptotic guarantees for our approximate soft-truncated likelihood in Section 3.
Establishing a convergence of our approximation in Equation 15 to the exact counterpart in Theorem \ref{asymptotic:guarantee}, our main result in Theorem \ref{multi:consistency:mle} proves a global consistency guarantee for the approximate selective MLE.

Suppose, we observe $\widehat{\mathrm{S}}_n =\mathrm{S}$; the subscript $n$ specifically indicates the dependence of $\widehat{\mathrm{S}}_n$ on the sample size.
In the asymptotic setting, let $$\sqrt{n}\widehat{\beta}_{\mathrm{S},n}\sim N(\sqrt{n} \beta_{\mathbb{M}_{\mathrm{S}}, \mathrm{S}, n}, \Sigma_{\mathbb{M}_{\mathrm{S}}, \mathrm{S}})$$
for any fixed value $\mathrm{S}$.
We denote our selective MLE by $\widehat{\beta}_{\mathbb{M}_{\mathrm{S}}, \mathrm{S}, n}^{\text{\;mle}}$.
Consider the following sequence of parameters 
$$\sqrt{n}\beta_{\mathbb{M}_{\mathrm{S}}, \mathrm{S}, n}= \bar{b}_n\beta_{\mathbb{M}_{\mathrm{S}}, \mathrm{S}}$$ such that $n^{-1/2}\bar{b}_n= O(1)$, $\bar{b}_n\to \infty$ as $n\to \infty$. 
Based on the K.K.T. conditions of optimality, our optimization variables $O_{1,n}$ are now defined as
\begin{equation}
\label{kkt:normal}
\sqrt{n}W_n = P_{\mathrm{S}}\sqrt{n}\widehat{\beta}_{\mathrm{S},n} + Q_{\mathrm{S}}\sqrt{n} O_{1,n} + r_{\mathrm{S}},
\end{equation}
where $\sqrt{n}W_n \sim N(0, \Sigma_{\mathbb{W}})$, and the selection event is equivalent to the linear constraints
\begin{equation}
\label{sel:region}
\sqrt{n} \ U  O_{1,n} < v.
\end{equation}
In a fixed $p$ and $n\to \infty$ regime, Theorem \ref{asymptotic:guarantee} verifies that the upper bound in Proposition 3.1 consistently approximates the exact selection probability in the following sense. 

\begin{theorem}
\label{asymptotic:guarantee}
Assume 
\begin{equation}
\label{assumption:asymp}
\begin{aligned}
& \frac{1}{(\bar{b}_n)^{2}} \Big\{\log\mathbb{P}\left[\sqrt{n} \ U  O_{1,n} < v\right]-\log\mathbb{P}\left[\sqrt{n} \ U  O_{1,n} < v_n\right]\Big\}
\end{aligned}
\end{equation}
converges to $0$ as $n\to \infty$, whenever $v_n=O(1)$.
Then, we have
\begin{equation*}
\begin{aligned}
&\lim_n \ \frac{1}{(\bar{b}_n)^{2}} \log\mathbb{P}\left[ \sqrt{n} \ U  O_{1,n} < v \right] \\
&\;+ \displaystyle\inf_{(\widetilde{\beta}_{\mathrm{S}}, o_1): \ Uo_1 \leq \frac{1}{\bar{b}_n}v}\;\Bigg\{ \dfrac{1}{2} \left(\widetilde{\beta}_{\mathrm{S}}-  J\beta_{\mathbb{M}_{\mathrm{\mathrm{S}}}, \mathrm{\mathrm{S}}} - \frac{1}{\bar{b}_n}k\right)^{\intercal}  \Sigma^{-1}\left(\widetilde{\beta}_{\mathrm{S}}- J\beta_{\mathbb{M}_{\mathrm{\mathrm{S}}}, \mathrm{\mathrm{S}}} - \frac{1}{\bar{b}_n}k\right) \\
&\;\;\;\;\;\;\;\;\;\;\;\;\;\;\;\;\;\;\;\;\;\;\;\;\;\;\;\;\;\;\;\;\;+  \dfrac{1}{2} \left(o_1  -A\widetilde{\beta}_{\mathrm{S}} -\frac{1}{\bar{b}_n}b\right)^{\intercal} \bar{\Sigma}^{-1}\left(o_1-A\widetilde{\beta}_{\mathrm{S}} -\frac{1}{\bar{b}_n}b\right)\Bigg\}=0.
\end{aligned}
\end{equation*}
\end{theorem}

Before we provide the proof, note, Theorem \ref{asymptotic:guarantee} suggests approximating $$\log\mathbb{P}\left[ \sqrt{n} \ U  O_{1,n} < v\right]$$ by
\begin{equation*}
\begin{aligned}
& -(\bar{b}_n)^2\cdot\displaystyle\inf_{(\widetilde{\beta}_{\mathrm{S}}, o_1): \bar{b}_n U o_1 \leq  v}\;\Bigg\{  \dfrac{1}{2} \left(\widetilde{\beta}_{\mathrm{S}}-  J\beta_{\mathbb{M}_{\mathrm{\mathrm{S}}}, \mathrm{\mathrm{S}}} - \frac{1}{\bar{b}_n}k\right)^{\intercal}  \Sigma^{-1}\left(\widetilde{\beta}_{\mathrm{S}}- J\beta_{\mathbb{M}_{\mathrm{\mathrm{S}}}, \mathrm{\mathrm{S}}} - \frac{1}{\bar{b}_n}k\right) \\
& \;\;\;\;\;\;\;\;\;\;\;\;\; \;\;\;\;\;\;\;\;\;\;\;\;\; \;\;\;\;\;\;\;\;\;+  \dfrac{1}{2} \left(o_1  -A\widetilde{\beta}_{\mathrm{S}} -\frac{1}{\bar{b}_n}b\right)^{\intercal} \bar{\Sigma}^{-1}\left(o_1-A\widetilde{\beta}_{\mathrm{S}} -\frac{1}{\bar{b}_n}b\right)\Bigg\}.
\end{aligned}
\end{equation*}
Observe, the display on the right-hand side is equivalent to
\begin{equation}
\label{our:approx}
\begin{aligned}
&-\displaystyle\inf_{(\widetilde{\beta}_{\mathrm{S}}, o_1): \sqrt{n}U o_1 \leq  v}\;\Big\{\dfrac{1}{2} (\sqrt{n}\widetilde{\beta}_{\mathrm{S}}-  J\sqrt{n}\beta_{\mathbb{M}_{\mathrm{\mathrm{S}}}, \mathrm{\mathrm{S}}} -k)^{\intercal}  \Sigma^{-1}(\sqrt{n}\widetilde{\beta}_{\mathrm{S}}- J\sqrt{n}\beta_{\mathbb{M}_{\mathrm{\mathrm{S}}}, \mathrm{\mathrm{S}}} - k) \\
& \;\;\;\;\;\;\;\;\;\;\;\;\; \;\;\;\;\;\;\;\;\;\;\;\;\; \;\;\;\;\;\;\;\;\;+  \dfrac{1}{2} ( \sqrt{n}o_1  -A\sqrt{n}\widetilde{\beta}_{\mathrm{S}} - b)^{\intercal} \bar{\Sigma}^{-1}(\sqrt{n}o_1-A\sqrt{n}\widetilde{\beta}_{\mathrm{S}} -  b)\Big\}.
\end{aligned}
\end{equation}
The approximate value in \eqref{our:approx} coincides with the bound in Proposition 3.1, as we highlight the dependence of our variables and the parameter vector on $n$ in our asymptotic setting. 

\begin{proof}
We write the variables $\widehat{\beta}_{\mathrm{S},n}$ and $O_{1,n} $ as a sum of i.i.d. variables
$$\begin{pmatrix} \widehat{\beta}_{\mathrm{S},n}\\ O_{1,n} \end{pmatrix}= n^{-1}\sum_{j=1}^n \begin{pmatrix} \widehat{\beta}^j_{\mathrm{S},n} \\ O^j_{1,n}\end{pmatrix},$$
where $\begin{pmatrix} (\widehat{\beta}^j_{\mathrm{S},n})^{\intercal} & (O^j_{1,n})^{\intercal}\end{pmatrix}^{\intercal}$ are distributed as Gaussian variables with
the covariance matrix:
$$ \Bigg(\begin{bmatrix}\Sigma^{-1} + A^{\intercal} \bar{\Sigma}^{-1} A & -A^{\intercal} \bar{\Sigma}^{-1} \\   -\bar{\Sigma}^{-1}A & \bar{\Sigma}^{-1} \end{bmatrix}\Bigg)^{-1},$$
and the mean vector:
 $$\Bigg(\begin{bmatrix}\Sigma^{-1} + A^{\intercal} \bar{\Sigma}^{-1} A & -A^{\intercal} \bar{\Sigma}^{-1} \\   -\bar{\Sigma}^{-1}A & \bar{\Sigma}^{-1} \end{bmatrix}\Bigg)^{-1}\begin{pmatrix}\Sigma^{-1}(J\beta_{\mathbb{M}_{\mathrm{\mathrm{S}}}, \mathrm{\mathrm{S}},n} + n^{-1/2} k) - n^{-1/2} A^{\intercal} \bar{\Sigma}^{-1}b \\ n^{-1/2}\bar{\Sigma}^{-1} b \end{pmatrix}.$$
 Marginally,  the covariance and mean for  $O^j_{1,n}$ are equal to
 $$\widetilde{\Sigma}= \bar{\Sigma}+ A\Sigma A^{\intercal},$$ and $$\widetilde{\mu}_n= n^{-1/2}A ( \bar{b}_n  J \beta_{\mathbb{M}_{\mathrm{\mathrm{S}}}, \mathrm{\mathrm{S}}} + k) + n^{-1/2}b,$$  
 respectively.
For the remaining proof, we let $O^{j; (c) }_{1,n}$ be the centered version of $O^{j}_{1,n}$.

We first examine the limit of
 $$\mathbb{P} \Big[ \frac{1}{\bar{b}_n} U \sum_{j=1}^n n^{-1/2}  O^{j; (c) }_{1,n} < - UA J\beta_{\mathbb{M}_{\mathrm{\mathrm{S}}}, \mathrm{\mathrm{S}}}\Big].$$
 Applying a large deviations limit, we have
 \begin{equation*}
 \label{eqn:inter}
\begin{aligned}
& \lim_n \ \frac{1}{(\bar{b}_n)^{2}} \log \mathbb{P} \Big[ \frac{1}{\bar{b}_n}U \sum_{j=1}^n n^{-1/2}  O^{j; (c) }_{1,n}< - UA J\beta_{\mathbb{M}_{\mathrm{\mathrm{S}}}, \mathrm{\mathrm{S}}} \Big]\\
&\Scale[0.95]{+ \displaystyle\inf_{o_1: Uo_1 \leq \frac{1}{\bar{b}_n}v}\;\; \dfrac{1}{2} \left(o_1  -A\left(J\beta_{\mathbb{M}_{\mathrm{\mathrm{S}}}, \mathrm{\mathrm{S}}}+\frac{1}{\bar{b}_n} k\right) -\frac{1}{\bar{b}_n} b\right)^{\intercal} \widetilde{\Sigma}^{-1}\left(o_1-A\left(J\beta_{\mathbb{M}_{\mathrm{\mathrm{S}}}, \mathrm{\mathrm{S}}}+\frac{1}{\bar{b}_n} k\right)-\frac{1}{\bar{b}_n} b\right)=0},
\end{aligned}
\end{equation*}
which can be rewritten as
\begin{equation}
\label{final:asymp}
\begin{aligned}
& \lim_n \frac{1}{(\bar{b}_n)^{2}} \log \mathbb{P} \Big[ \frac{1}{\bar{b}_n}U \sum_{j=1}^n n^{-1/2}  O^{j; (c) }_{1,n} < - UA J\beta_{\mathbb{M}_{\mathrm{\mathrm{S}}}, \mathrm{\mathrm{S}}} \Big]\\
&\;\;\;\;\;\;\;+\displaystyle\inf_{(\widetilde{\beta}_{\mathrm{S}}, o_1): Uo_1 \leq \frac{1}{\bar{b}_n}v}\;\Bigg\{ \dfrac{1}{2} \left(\widetilde{\beta}_{\mathrm{S}}-  J\beta_{\mathbb{M}_{\mathrm{\mathrm{S}}}, \mathrm{\mathrm{S}}} - \frac{1}{\bar{b}_n} k\right)^{\intercal}  \Sigma^{-1}(\widetilde{\beta}_{\mathrm{S}}- J\beta_{\mathbb{M}_{\mathrm{\mathrm{S}}}, \mathrm{\mathrm{S}}} - \frac{1}{\bar{b}_n} k) \\
&\;\;\;\;\;\;\;\;\;\;\;\;\;\;\;\;\;\;\;\;\;\;\;\;\;\;\;\;\;\;\;+  \dfrac{1}{2} \left(o_1  -A\widetilde{\beta}_{\mathrm{S}} -\frac{1}{\bar{b}_n}b\right)^{\intercal} \bar{\Sigma}^{-1}\left(o_1-A\widetilde{\beta}_{\mathrm{S}} -\frac{1}{\bar{b}_n}b\right)\Bigg\}=0.
\end{aligned}
\end{equation}

\noindent Completing our proof, observe, the probability of selection simplifies as
\begin{equation*}
\begin{aligned}
 \mathbb{P} \left[ \sqrt{n} \ U O_{1,n}< v \right]
&= \mathbb{P} \left[ U\sum_{j=1}^n n^{-1/2}  O^{j; (c) }_{1,n}  < v - U A ( \bar{b}_n  J \beta_{\mathbb{M}_{\mathrm{\mathrm{S}}}, \mathrm{\mathrm{S}}} + k) - U b\right]\\
&= \mathbb{P} \left[\frac{1}{\bar{b}_n} U \sum_{j=1}^n n^{-1/2}  O^{j; (c) }_{1,n}< \frac{1}{\bar{b}_n}(v - \bar{b}_nU A J \beta_{\mathbb{M}_{\mathrm{\mathrm{S}}}, \mathrm{\mathrm{S}}}-Uk -Ub)\right].
\end{aligned}
\end{equation*}
Using our assumption in \eqref{assumption:asymp}, we have
\begin{equation*}
\begin{aligned}
& \lim_n \ \frac{1}{(\bar{b}_n)^{2}} \Big\{\log \mathbb{P} \left[ \sqrt{n} \ U O_{1,n}< v\right]-\log\mathbb{P} \Big[ \frac{1}{\bar{b}_n} U \sum_{j=1}^n n^{-1/2} O^{j; (c) }_{1,n}< - UA J\beta_{\mathbb{M}_{\mathrm{\mathrm{S}}}, \mathrm{\mathrm{S}}} \Big]\Big\}=0,
\end{aligned}
\end{equation*}
which proves the claim.
\end{proof}

Let $\mathcal{B}_{U; v}(\cdot)$ be a barrier penalty such that
$(\bar{b}_n)^{-2}\mathcal{B}_{U; v}(\bar{b}_n o_1)$ converges to $0$ in a pointwise sense
only if $Uo_1 \leq \frac{1}{\bar{b}_n}v$, and assumes the value infinity otherwise. 
Based on Theorem \ref{asymptotic:guarantee}, the second term in the limit trivially agrees with 
\begin{equation}
\label{barrier:penalty}
\begin{aligned}
&\displaystyle\inf_{\widetilde{\beta}_{\mathrm{S}}, o_1}\;\Bigg\{  \dfrac{1}{2} \left(\widetilde{\beta}_{\mathrm{S}}-  J\beta_{\mathbb{M}_{\mathrm{\mathrm{S}}}, \mathrm{\mathrm{S}}} - \frac{1}{\bar{b}_n}  k\right)^{\intercal}  \Sigma^{-1}\left(\widetilde{\beta}_{\mathrm{S}}- J\beta_{\mathbb{M}_{\mathrm{\mathrm{S}}}, \mathrm{\mathrm{S}}} - \frac{1}{\bar{b}_n}  k\right) \\
&\;\;\;\;\;\;\;\;\;+  \dfrac{1}{2} \left(o_1  -A\widetilde{\beta}_{\mathrm{S}} -\frac{1}{\bar{b}_n}  b\right)^{\intercal} \bar{\Sigma}^{-1}\left(o_1-A\widetilde{\beta}_{\mathrm{S}} -\frac{1}{\bar{b}_n}  b\right) + \frac{1}{(\bar{b}_n)^{2}}\mathcal{B}_{U; v}(\bar{b}_n o_1) \Bigg\}
\end{aligned}
\end{equation}
as $n\to \infty$.

\begin{theorem}
 \label{multi:consistency:mle}
Fix $\delta>0$. Consider the assumptions in Theorem \ref{asymptotic:guarantee}. 
Then
$$\mathbb{P}\left[\|\widehat{\beta}_{\mathbb{M}_{\mathrm{S}}, \mathrm{S}, n}^{\text{\;mle}}- \beta_{\mathbb{M}_{\mathrm{S}}, \mathrm{S}, n}\|_2 >\delta  \; \lvert \; \widehat{\mathrm{S}}_n =\mathrm{S},\; \widehat{\beta}^{\perp}_{\mathrm{S},n}= \widehat{\beta}^{\perp}_{\mathrm{S}} \right] \to 0$$
as $n\to\infty$.
\end{theorem}
\medskip

\begin{proof}
Let $\eta_{\mathbb{M}_{\mathrm{S}}, \mathrm{S},n} = \Sigma^{-1}\left(J \beta_{\mathbb{M}_{\mathrm{\mathrm{S}}}, \mathrm{\mathrm{S}}} + \frac{1}{\bar{b}_n}  k\right)$, let $\alpha_n(\eta_{\mathbb{M}_{\mathrm{S}}, \mathrm{S},n})$ be given by
\begin{equation*}
\begin{aligned}
&  \frac{1}{2} \eta^{\intercal}_{\mathbb{M}_{\mathrm{S}}, \mathrm{S},n} \Sigma  \eta_{\mathbb{M}_{\mathrm{S}}, \mathrm{S},n} - \displaystyle\inf_{\widetilde{\beta}_{\mathrm{S}}, o_1}\;\Big\{ \dfrac{1}{2} (\widetilde{\beta}_{\mathrm{S}}- \Sigma\eta_{\mathbb{M}_{\mathrm{S}}, \mathrm{S},n})^{\intercal} \Sigma^{-1}(\widetilde{\beta}_{\mathrm{S}}-  \Sigma\eta_{\mathbb{M}_{\mathrm{S}}, \mathrm{S},n}) \\
& \;\;\;\;\;\;\;\;\;\;\;\;\; \;\;\;\;\;\;\;\;\;\;\;\;\; +  \dfrac{1}{2} \left( o_1  -A\widetilde{\beta}_{\mathrm{S}} - \frac{1}{\bar{b}_n} b\right)^{\intercal} \bar{\Sigma}^{-1}\left(o_1-A\widetilde{\beta}_{\mathrm{S}} -  \frac{1}{\bar{b}_n} b\right) + \frac{1}{(\bar{b}_n)^{2}}\mathcal{B}_{U; v}(\bar{b}_n o_1)\Big\}.
\end{aligned}
\end{equation*}
and let $\bar{\alpha}_n(\eta_{\mathbb{M}_{\mathrm{S}}, \mathrm{S},n})$ assume the value
$$ \frac{1}{2} \eta^{\intercal}_{\mathbb{M}_{\mathrm{S}}, \mathrm{S},n} \Sigma \eta_{\mathbb{M}_{\mathrm{S}}, \mathrm{S},n} + \frac{1}{(\bar{b}_n)^{2}} \log\mathbb{P}\left[ \sqrt{n} \ U  O_{1,n} < v\right].$$
To simplify notations in the proof, let us denote the conditional expectation
$$\mathbb{\mathrm{E}}\left[\cdot\; \lvert\; \widehat{\mathrm{S}}_n =\mathrm{S},\; \widehat{\beta}^{\perp}_{\mathrm{S},n}= \widehat{\beta}^{\perp}_{\mathrm{S}}\right]$$ by $\overline{\mathbb{\mathrm{E}}}[\cdot]$.
Using the finite sample bound in Theorem 3, we have
\begin{equation*}
\begin{aligned}
 \frac{1}{(\bar{b}_n)^{2}}\ \overline{\mathbb{\mathrm{E}}}\left[n\|\widehat{\beta}_{\mathbb{M}_{\mathrm{S}}, \mathrm{S}, n}^{\text{\;mle}}- \beta_{\mathbb{M}_{\mathrm{S}}, \mathrm{S}, n}\|_2^2 \right] 
& \leq (B)^{-1}\overline{\mathbb{\mathrm{E}}}\left[\|\sqrt{n}(\bar{b}_n)^{-1}\ \widehat{\beta}_{\mathrm{S}, n}- \grad \alpha_n(\eta_{\mathbb{M}_{\mathrm{S}}, \mathrm{S},n})\|_2^2\right]\\
&= (B)^{-1} \overline{\mathbb{\mathrm{E}}}\left[\|\sqrt{n}(\bar{b}_n)^{-1}\ \widehat{\beta}_{\mathrm{S}, n}- \grad \alpha_n(\eta_{\mathbb{M}_{\mathrm{S}}, \mathrm{S},n})\|_2^2\right]\\
&\leq  (B)^{-1}\frac{1}{(\bar{b}_n)^{2}}\overline{\mathbb{\mathrm{E}}}\left[\|\sqrt{n}\widehat{\beta}_{\mathrm{S}, n} - \bar{b}_n\grad \bar\alpha_n(\eta_{\mathbb{M}_{\mathrm{S}}, \mathrm{S},n})\|_2^2\right]\\
&+ (B)^{-1} \|\grad \bar\alpha_n(\eta_{\mathbb{M}_{\mathrm{S}}, \mathrm{S},n})- \grad \alpha_n(\eta_{\mathbb{M}_{\mathrm{S}}, \mathrm{S},n})\|_2^2.
\end{aligned}
\end{equation*}
The last display is obtained after using a triangle inequality. 
The first term on the right hand side is
$$(B)^{-1}\frac{1}{(\bar{b}_n)^{2}}\ \text{Var}\left( \sqrt{n}\widehat{\beta}_{\mathrm{S}, n}  \lvert  \; \widehat{\mathrm{S}}_n =\mathrm{S},\; \widehat{\beta}^{\perp}_{\mathrm{S},n}= \widehat{\beta}^{\perp}_{\mathrm{S}}\right).$$
Applying the variance reduction result in \cite{kanter1977reduction}, this term is further bounded by 
$$(B)^{-1}\frac{1}{(\bar{b}_n)^{2}}\ \text{Var}( \sqrt{n}\widehat{\beta}_{\mathrm{S}, n}),$$
and converges to $0$ as $n\to \infty$. 
The convergence of the second term uses the convexity and differentiability of the functions $\alpha_n$ and $\bar\alpha_n(\cdot)$ in addition to the fact
$$\bar\alpha_n(\eta_{\mathbb{M}_{\mathrm{S}}, \mathrm{S},n})-\alpha_n(\eta_{\mathbb{M}_{\mathrm{S}}, \mathrm{S},n}) \to 0$$ as $n\to \infty$.
These facts imply
$$\lim_n \frac{1}{(\bar{b}_n)^{2}}\ \overline{\mathbb{\mathrm{E}}}\left[n\|\widehat{\beta}_{\mathbb{M}_{\mathrm{S}}, \mathrm{S}, n}^{\text{\;mle}}- \beta_{\mathbb{M}_{\mathrm{S}}, \mathrm{S}, n}\|_2^2 \right] =0.$$
Our claim follows immediately by observing
\begin{equation*}
\begin{aligned}
& \mathbb{P}\left[\frac{1}{\bar{b}_n}\sqrt{n}\ \|\widehat{\beta}_{\mathbb{M}_{\mathrm{S}}, \mathrm{S}, n}^{\text{\;mle}}- \beta_{\mathbb{M}_{\mathrm{S}}, \mathrm{S}, n}\|_2 >\delta \; \lvert \; \widehat{\mathrm{S}}_n =\mathrm{S},\; \widehat{\beta}^{\perp}_{\mathrm{S},n}= \widehat{\beta}^{\perp}_{\mathrm{S}}\right]\\
&\;\;\;\;\;\;\;\;\;\;\;\;\;\;\;\;\;\;\;\;\;\;\;\;\;\;\;\;\;\;\;\;\;\;\;\;\;\;\;\;\;\;\;\;\; \leq \frac{1}{(\delta \cdot \bar{b}_n)^{2}}\ \overline{\mathbb{\mathrm{E}}}\left[n\|\widehat{\beta}_{\mathbb{M}_{\mathrm{S}}, \mathrm{S}, n}^{\text{\;mle}}- \beta_{\mathbb{M}_{\mathrm{S}}, \mathrm{S}, n}\|_2^2 \right] .
\end{aligned}
\end{equation*}
\end{proof}

\bibliographystyle{chicago}
\bibliography{references.bib}

\end{document}